\documentclass[acmsmall]{acmart}

\usepackage{array}
\newcolumntype{P}[1]{>{\centering\arraybackslash}p{#1}}
\usepackage{relsize}

\usepackage{listings}
\usepackage{xcolor} 
\usepackage{multirow}
\usepackage{booktabs} 
\usepackage{float}
\usepackage{amsmath}
\usepackage{textcomp}
\usepackage{graphicx}
\usepackage[belowskip=-1pt]{caption}
\usepackage{subcaption}
\usepackage{rotating}
\usepackage{tabularx}
\usepackage[noend,linesnumbered, ruled,vlined]{algorithm2e}
\usepackage{wrapfig}
\usepackage{listings}
\usepackage{color, colortbl}
\usepackage{balance} 
\usepackage{lscape}
\SetKw{KwStep}{Step}
\SetKw{KwIn}{in}
\usepackage{hyperref}
\usepackage{xspace}
\usepackage{framed}
\usepackage{syntax}
\usepackage{comment}
\usepackage{adjustbox}
\usepackage[flushleft]{threeparttable}
\newcolumntype{C}[1]{>{\centering\arraybackslash}p{#1}}
\usepackage{bbding}
\usepackage{pifont}
\usepackage{wasysym}
\usepackage{tcolorbox}
\usepackage[framemethod=TikZ]{mdframed}
\usepackage{soul}
\usepackage{flushend}
\usepackage{enumitem}

\usepackage{tikz}

\definecolor{codegreen}{rgb}{0,0.6,0}
\definecolor{codegray}{rgb}{0.5,0.5,0.5}
\definecolor{codepurple}{rgb}{0.58,0,0.82}
\definecolor{backcolour}{rgb}{0.95,0.95,0.92}
\definecolor{Gray}{gray}{0.1}
\lstdefinestyle{mystyle}{
	backgroundcolor=\color{backcolour},   
	commentstyle=\color{codegreen},
	keywordstyle=\color{magenta},
	numberstyle=\tiny\color{codegray},
	stringstyle=\color{codepurple},
	basicstyle=\scriptsize,
	breakatwhitespace=false,         
	breaklines=true,                 
	captionpos=b,                    
	keepspaces=true,                 
	numbers=left,                    
	numbersep=5pt,                  
	showspaces=false,                
	showstringspaces=false,
	showtabs=false,                  
	tabsize=2
}

\lstdefinelanguage{Pythonna}{%
	language     = python,
	morekeywords = {to_categorical, flow_from_directory, pad_sequences, load_image}
}

\lstdefinestyle{customc}{
	belowcaptionskip=1\baselineskip,
	breaklines=false,
	frame= single,
	breaklines = true,
	xleftmargin=\parindent,
	language= Pythonna,
	showstringspaces=false,
	basicstyle=\footnotesize\ttfamily,
	keywordstyle=\bfseries\color{green!40!black},
	commentstyle=\itshape\color{purple!40!black},
	identifierstyle=\color{blue},
	stringstyle=\color{codegreen},
	backgroundcolor=\color{gray!4}
}

\graphicspath{{./figures/}}

\newcommand{\secref}[1]{\S\ref{#1}}

\usepackage{lipsum}
\usepackage[htt]{hyphenat}
\newcommand{\ie}{\textit{i.e.,}\xspace}
\newcommand{\eg}{\textit{e.g.,}\xspace}
\newcommand{\etc}{\textit{etc.,}\xspace}

\newcommand{\via}{\textit{via}\xspace}
\newcommand{\etal}{\textit{et al.}\xspace}

\newcommand{\ourapproach}{Theia\xspace}
\newcommand{\nlint}{NeuraLint\xspace}

\newcommand{\sof}{\textit{Stack Overflow}\xspace}
\newcommand{\Keras}{\textit{Keras}\xspace}
\newcommand{\PyTorch}{\textit{PyTorch}\xspace}
\newcommand{\Caffe}{\textit{Caffe}\xspace}
\newcommand{\Theano}{\textit{Theano}\xspace}

\newcommand{\tensor}{\textit{Tensorflow}\xspace}
\newcommand{\gh}{\textit{GitHub}\xspace}

\newcounter{rqs}
\stepcounter{rqs}

\newcounter{NumObservations}
\stepcounter{NumObservations}
\definecolor{shadecolor}{rgb}{.9,.9,.9}

\AtBeginDocument{%
  \providecommand\BibTeX{{%
    \normalfont B\kern-0.5em{\scshape i\kern-0.25em b}\kern-0.8em\TeX}}}

\setcopyright{acmcopyright}
\copyrightyear{2023}
\acmYear{2023}
\acmDOI{XXXXXXX.XXXXXXX}

\acmJournal{JACM}
\acmVolume{37}
\acmNumber{4}
\acmArticle{111}
\acmMonth{8}

\begin{document}

\title {Leveraging Data Characteristics for Bug Localization in Deep Learning Programs }

\author{Ruchira Manke}
\email{rmanke@tulane.edu}
\affiliation{%
	\institution{Department of Computer Science, Tulane University}
	\streetaddress{303 Stanley Thomas Hall, New Orleans}
	\state{Louisiana}
	\postcode{70118}
	\country{USA}
}

\author{Mohammad Wardat}
\email{wardat@oakland.edu}
\affiliation{%
	\institution{Department of Computer Science and Engineering, Oakland University}
	\streetaddress{115 Library Drive}
	\city{ Rochester}
	\state{MI}
	\postcode{48309}
	\country{USA}
}

\author{Foutse Khomh}
\email{foutse.khomh@polymtl.ca}
\affiliation{%
	\institution{SWAT Lab., Polytechnique Montréal}
        \streetaddress{P.O. Box 6079, Station Centre- Ville}
	\state{Montréal}
        \postcode{H3C 3A7}
	\country{CA}
}

\author{Hridesh Rajan}
\email{hrajan@tulane.edu}
\affiliation{%
	\institution{School of Science and Engineering, Tulane University}
	\streetaddress{201 Lindy Boggs Center, New Orleans}
	\state{Louisiana}
	\postcode{70118}
	\country{USA}
}

\begin{abstract}
Deep Learning (DL) is a class of machine learning algorithms that are used in a wide variety of applications. 
Like any software system, DL programs can have bugs.
To support bug localization in DL programs, several tools have been proposed in the past.
As most of the bugs that occur due to 
improper model structure known as structural bugs
lead to inadequate performance during training,
it is challenging for developers to identify the root cause and address these bugs.
To support bug detection and localization in DL programs, in
this paper, we propose \ourapproach, which detects and localizes structural bugs in DL programs. 
Unlike the previous works, \ourapproach considers the training dataset characteristics to automatically detect bugs in DL programs developed using two deep learning libraries, \Keras and \PyTorch.
Since training the DL models is a time-consuming process, \ourapproach detects these bugs at the beginning of the training process  
and alerts the developer with informative messages containing the bug's location and actionable fixes which will help them to improve the structure of the model.
We evaluated \ourapproach on a benchmark of 40 real-world buggy  DL programs obtained from \sof. 
Our results show that \ourapproach successfully localizes 57/75 structural bugs in 40 buggy programs, 
whereas \nlint, a state-of-the-art approach capable of localizing structural bugs before training localizes 17/75 bugs.

\end{abstract}



\maketitle
\section{Introduction}\label{sec:intro}
Deep learning (DL) based software has recently gained popularity and is being used in various fields, including chatbots~\cite{Csaky19}, virtual assistants~\cite{Iannizzotto18}, and financial institutions~\cite{Roy18}.
Their popularity has drawn the interest of the software engineering community to understand their development process.
As bugs are inherent to the software development process, several studies have been conducted in the past to understand the characteristics of DL bugs, their root causes, and repair solutions~\cite{zhang18,islam19,humbatova20taxonomy,islam20repairing}.
To support the development of DL programs, several DL libraries and frameworks, such as Tensorflow~\cite{Tensorflow}, Keras~\cite{Keras}, and Pytorch~\cite{Pytorch} are available which provide various APIs for building, training, and evaluating these programs.
As DL programs are based on tensor operations, which are multi-dimensional arrays that generalize matrices to higher dimensions,
various operations on tensors, such as matrix multiplication, addition, activation functions, and convolutions are performed to build and train these models.
These libraries validate the correctness of the computations and use assertions to detect crash bugs.
However, the dependency of DL programs on data makes it challenging to impose assertions for silent bugs, that occur due to hidden logic errors and commonly lead to incorrect model behavior, inaccurate predictions, or degraded performance.
Although, these libraries provide \textit{`callbacks'} to monitor and customize various stages of training loops (\eg at the start or end of an epoch, at the start or end of the batch, \textit{etc.}), these callback methods (\eg EarlyStopping(), TerminateOnNaN()) 
do not indicate which layer or hyper-parameter caused the issue.
As a result, DL libraries lack comprehensive debugging mechanisms for locating silent bugs.
Researchers in the past~\cite{islam19,humbatova20taxonomy} have found that silent bugs are more prevalent ($>60\%$) than crash bugs in DL programs. In the software engineering community, these bugs are referred to as \textit{structural bugs} \cite{Ma18mode}.
In this work, we focus on structural bugs, which primarily arise from misconfigured hyper-parameters in the model.

To assist developers in identifying the structural bugs in DL programs, several techniques such as UMLAUT~\cite{schoop2021umlaut}, DeepLocalize~\cite{wardat21DeepLocalize}, DeepDiagnosis~\cite{wardat22DeepDiagnosis}, TheDeepChecker~\cite{BraiekDeepCheck}, DeepFD~\cite{cao22deepfd} for detecting and localizing these bugs have been proposed in the past.
These techniques observe the abnormal behavior during the training of the model and
identify structural bugs based on certain symptoms.
Due to one-to-many mapping between the abnormal behavior observed during training and its root causes~\cite{wardat22DeepDiagnosis}, these techniques are unable to provide sufficient insight into the underlying cause of the issue, thereby requiring several rounds to fix the bugs.
As training a DL model is expensive, identifying model inefficiencies during training wastes computational resources.
To overcome this problem, Nikanjam~\etal~\cite{nikanjam2021neuralint} proposed a static approach, \nlint, that examines the DL model for structural errors and design inefficiencies and can detect bugs in DL programs that are not covered by previous dynamic approaches~\cite{wardat21DeepLocalize,schoop2021umlaut,wardat22DeepDiagnosis,cao22deepfd}.

The current approaches for detecting and localizing bugs in DL programs
are either specifically designed for classification tasks~\cite{schoop2021umlaut,Zhang21Autotrainer}
or focused on identifying structural bugs that are common across different DL architectures (\ie Fully-Connected Neural Networks (FCNN), Convolutional Neural Networks (CNN), and Recurrent Neural Networks (RNN))~\cite{wardat21DeepLocalize,wardat22DeepDiagnosis,cao22deepfd,BraiekDeepCheck}.
\nlint~\cite{nikanjam2021neuralint} covers CNN architecture-specific structural bugs. However, it relies on the parsed source code of the DL model and does not consider the training data to localize the bugs, resulting in false alarms.
As the DL models are data-driven, information acquired using only the parsed source code is insufficient for effective bug localization.
Our insight is that, as the DL models are data-driven, combining the training data characteristics with the model’s source code provides a more comprehensive analysis, improving the bug localization accuracy.

\subsection{Motivation}

\begin{figure}[h]
	\centering
     \begin{subfigure}[t]{0.48\textwidth}
         \includegraphics[width=\textwidth]{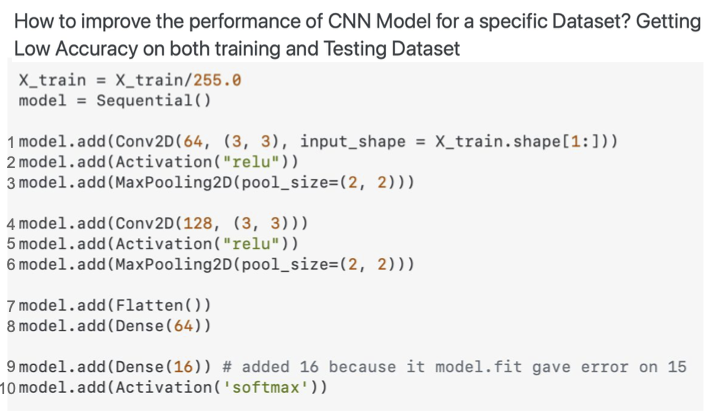}
     \end{subfigure}
     \hfill
     \begin{subfigure}[t]{0.48\textwidth}
         \includegraphics[width=\textwidth]{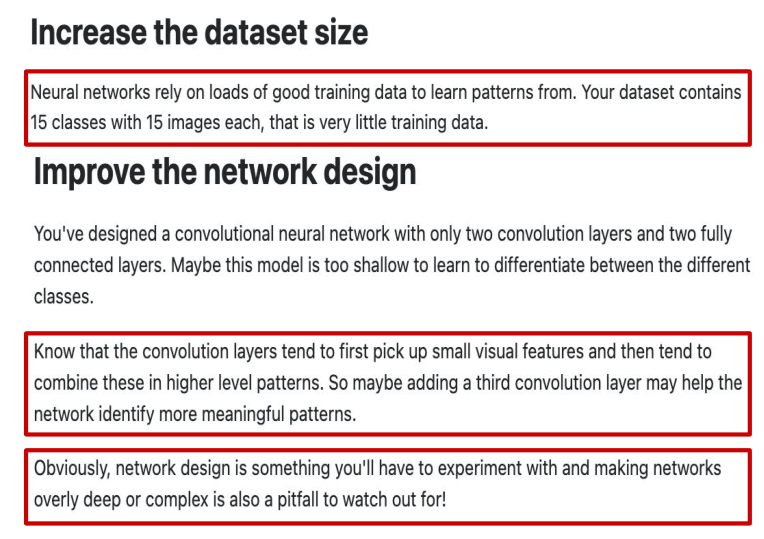}
     \end{subfigure}
    
    \caption{An example from \sof~\cite{SO9} with structural bug. \\}
	\label{motivating}
\end{figure}

In practice, new developers usually use familiar solutions when
designing the DL program without fully understanding the effect of those
solutions and various factors that need to be considered for their application. For instance,
utilizing the architecture of the model designed for multi-class classification for binary classification or a simple
model designed for gray-scale image datasets for more complex color images leads to inadequate performance during training.
Due to the numerous hyper-parameters in DL models, debugging these programs is challenging~\cite{islam20repairing,humbatova20taxonomy}.
For instance, Fig.~\ref{motivating} presents a query posted on \sof~\cite{SO9}, wherein the developer has implemented a CNN program using \Keras.
The program demonstrated erratic behavior during training and testing.
The developer in the post mentions the various CNN architectures that were attempted but did not achieve the desired results. In response, another \sof user 
pointed out the following issues in the CNN program. 
First, the dataset used to train the model is very small, and the user
recommends increasing the dataset size by adding high-quality data.
Secondly, as the model is designed for multi-class classification, the user also suggests improving the network design by adding more convolution layers and explaining the impact of shallow and overly deep networks on performance.
As a side note, the user also suggests lowering the learning rate.
The developer could not design a CNN program with good performance and locate the cause of these bugs because structural bugs usually affect the program's performance rather than causing the program to crash. 

For the CNN program in Fig.~\ref{motivating}, current state-of-the-art techniques~\cite{schoop2021umlaut, wardat21DeepLocalize,wardat22DeepDiagnosis,Zhang21Autotrainer,cao22deepfd, BraiekDeepCheck} were not able to identify structural flaws, as mentioned in the \sof post.
These techniques primarily focus on identifying bugs using different parameters such as weights, gradients, loss, and accuracy within the designed model, assuming that the model’s depth and width are appropriately defined. Therefore, these techniques do not identify structural flaws due to suboptimal model depth or width, which can significantly affect performance, as shown in Fig.~\ref{motivating}.
While \nlint identifies structural errors before training, as it relies on the model’s parsed source code and does not capture the characteristics of the training dataset in the meta-model, it cannot determine whether the model is too shallow or narrow for the training dataset. For the example in Fig.~\ref{motivating}, existing approaches cannot identify the structural bug due to the insufficient number of layers in the model. 
Choosing an appropriate number of layers or neurons during model design is challenging. In practice, it is usually done by manually fine-tuning the model or using automated tools like Auto-Keras~\cite{jin2019auto}.
However, fine-tuning is expensive~\cite{ippolito2022hyperparameter}; on high-performance machines, Auto-Keras usually requires 8-12 hours to search for models with reasonable accuracy (90\% or higher)~\cite{jin2019auto}.
Developers often need to pay more attention to the fundamental design principles,
which leads to incorrect model behavior during training and requires 
significant debugging time and effort. 
Therefore, some lightweight automated debugging tools are needed to verify the designed model structure aligns with the training dataset and task before initiating an expensive training process.

\subsection{Contributions}
In this paper, we propose a technique, named \ourapproach, 
which leverages the characteristics of the training dataset along with the model's parsed source code for localizing the structural bugs in DL programs, \ie bugs related
to the activation function, layer properties, model properties, loss function, preprocessing of data, and
bugs due to missing/redundant/wrong layers 
and provide suggestions to fix the bug.
These bugs lead to performance issues, \ie low/stuck accuracy during training.
Therefore, the scope of \ourapproach is to localize structural bugs in DL programs.
\ourapproach supports two types of DL architecture, \ie Fully-Connected Neural Networks (FCNNs) and Convolutional Neural Networks (CNNs) designed for regression, as well as classification tasks. 

To design \ourapproach, first, a general representation of the DL program, a \textit{meta-model}, that is independent of any DL libraries or frameworks is constructed. 
A meta-model captures the characteristics of the dataset, \eg dimension, type of training data, and properties of the DL model, \eg the number of convolution layers and learning rate.
\ourapproach utilizes the meta-model and performs context-sensitive analysis, namely call-strings analysis~\cite{Sharir78} and parameter-sensitive analysis~\cite{Wei15} using the verification rules to detect the structural bugs in DL programs developed using two popular deep learning libraries, \Keras and \PyTorch.
\ourapproach detects these bugs at the beginning of the training process and alerts the developer with informative messages that include the bug's location and fix recommendations to improve the structure of the DL model.

We evaluated \ourapproach on 40 real-world buggy DL programs obtained from \sof designed for regression and classification tasks. 
\ourapproach successfully finds 57/75 bugs and is more
effective than \nlint~\cite{nikanjam2021neuralint}, which detects 17/75 bugs.

In summary, this paper makes the following contributions:
\begin{itemize}
    \item We investigated the mapping between the characteristics of the dataset and the structure of the model.
    \item We provide verification rules to detect the occurrence of structural bugs.
    \item We designed and implemented \ourapproach, for two popular DL libraries, \Keras and \PyTorch, 
    for automatically detecting structural bugs at the beginning of the training process.
    \item We evaluated \ourapproach on 40 buggy DL programs and compared with \nlint~\cite{nikanjam2021neuralint}. 
    We found that \ourapproach is more effective and efficient compared to \nlint which can be used by developers to detect structural bugs in DL programs.
   
\end{itemize}

The rest of the paper is organized as follows.
\secref{sec:background} describes the background.
\secref{sec:Study} describes the deep learning program structural bugs.
\secref{sec:Approach} describes the verification rules, explains how they are used to detect structural bugs in our approach, and presents an algorithm for identifying these bugs.
\secref{sec:evaluation} describes the evaluation of our approach compared with prior work.
\secref{sec:threatstovalidity} discusses the threats to validity.
\secref{sec:relatedwork} discusses related work, 
\secref{sec:conclusion} concludes 
and discusses future work, and
\secref{sec:data-availability} provides details of the replication package.

\section{Background}\label{sec:background}
\subsection{Deep Learning Programs}
Deep learning has recently been widely used in different domains to automatically learn complex patterns from data~\cite{cao22deepfd}. Deep learning architectures (\ie FCNN, CNN) comprise many layers with each layer serving a distinct function.
These layers are fundamental building blocks that transform input data into meaningful output.
For example, the FCNN program comprises an input layer followed by a series of fully connected layers that learn features from input, and finally an output layer that is trained to predict the output. 
CNN program has a more complex structure comprising convolution and pooling layers followed by fully connected layers. Convolution layers extract the features from the input and produce feature maps, pooling layers help in reducing the size of feature maps, and fully connected layers help the model learn class-wise features. Using features extracted from previous layers, the output layer is trained to predict the final output.
The DL program has two types of parameters: (1) the model
parameters that are learned during training; and (2) the
hyper-parameters whose value can be configured before training, \eg number of neurons, filters in the convolution
layer, kernel size, strides.
Each layer has a different number of hyper-parameters that help the model to learn and are provided by the developer while designing the DL programs.

\subsection{Deep Learning Library }
\Keras and \PyTorch are the popular deep learning libraries that provide APIs for implementing different stages of a DL program, namely, data preparation, modeling, and training~\cite{biswas22art}. 
The APIs are written in the form of classes, \eg \texttt{Conv2D}, \texttt{MaxPool2d}, \texttt{Dense}, \etc comprising many methods with parameters.
Some of these parameters are used in structuring the model, while others serve as \textit{hyper-parameter}s for the DL program which helps in learning.
The hyper-parameters are initialized by the developer while using the API and must be configured considering the task for which the DL program is designed as well as the characteristics of the training dataset.
Failing to consider them while designing the DL programs results in the incorrect configuration of hyper-parameters which may not necessarily cause a program to crash, however, results in a program with performance issues, \eg incorrect output or low/stuck classification accuracy.

\section{Deep Learning Program Bugs} \label{sec:Study}
In this section, we first describe the process utilized to identify the structural bugs that lead to performance issues. Then, we discuss the mapping between the characteristics of the dataset and the structural bug. Followed by verification rule creation methodology.

\begin{figure}[t!]
    \centering
    \subfloat[\centering Distribution of different types of DL programs with varying depths]{\includegraphics[width=6cm]{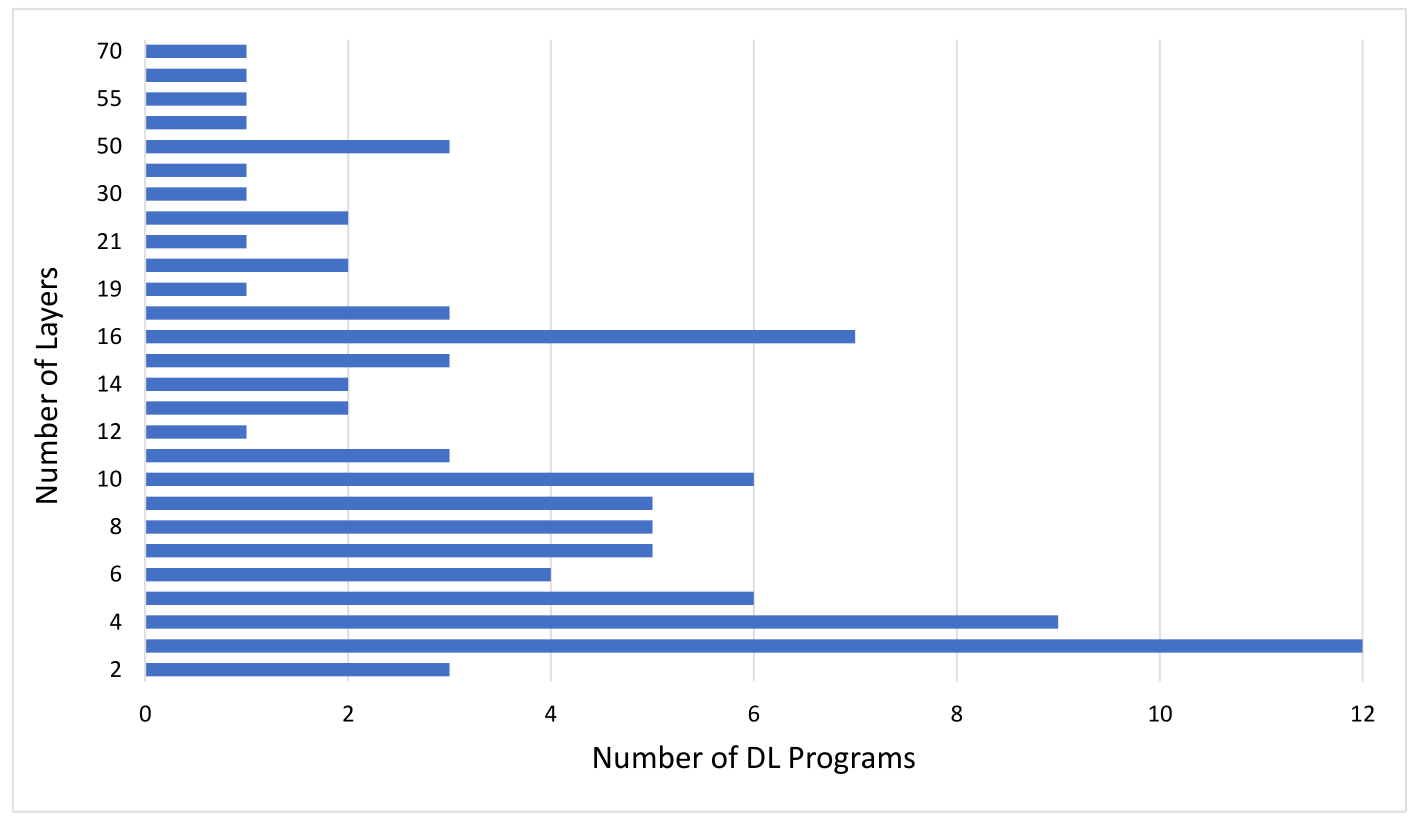}}%
    \qquad
    \subfloat[\centering Distribution of bugs in different categories]{\includegraphics[width=6cm]{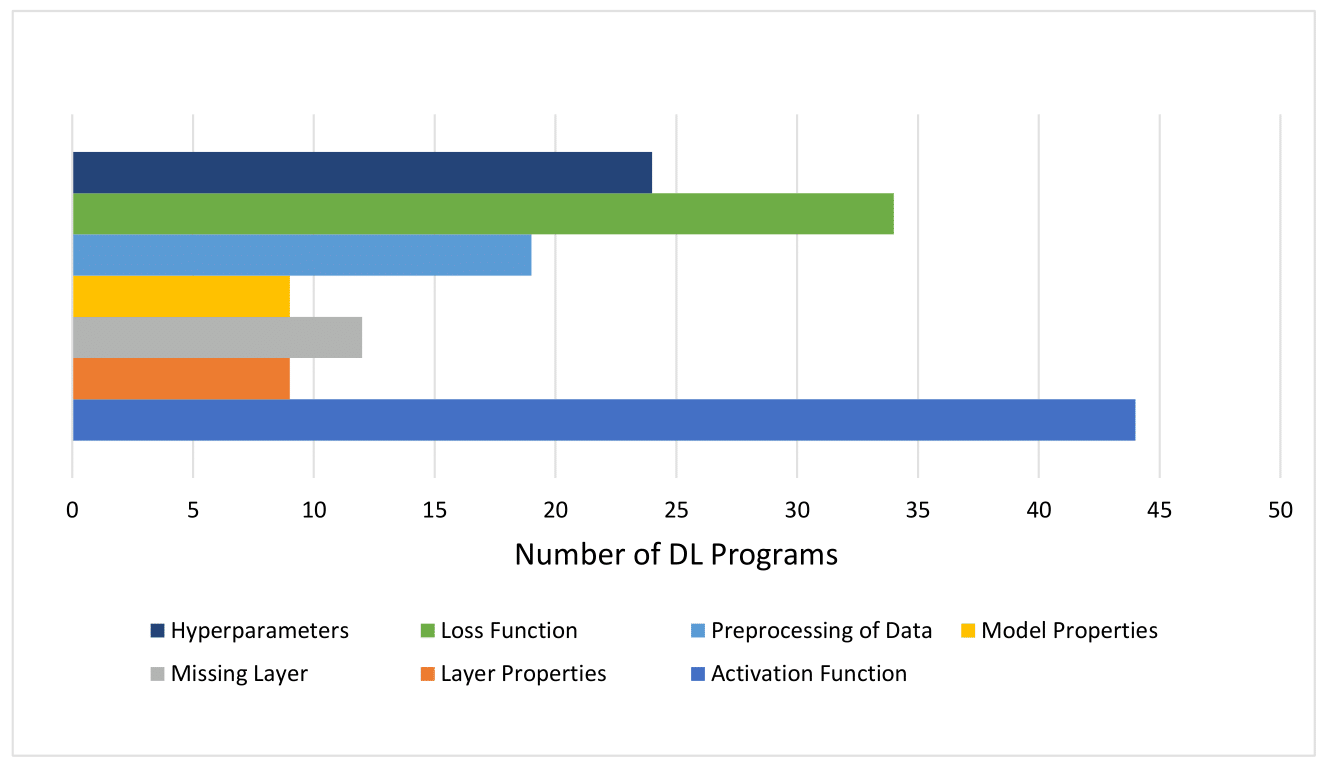}}%
    \caption{Details of DL programs used for mapping.}%
    \label{ModelLayer}
\end{figure}

\subsection{Structural Bugs Investigation}
Researchers in the past have studied deep learning program bugs, and their characteristics and also provided a taxonomy of faults for these programs.
Zhang~\etal~\cite{zhang18} studied the root cause of bugs and their symptoms in TensorFlow programs.
This research was extended by Islam~\etal~\cite{islam19} and they studied the types of bugs, their root causes, and their impacts using five popular DL libraries. 
Humbatova~\etal~\cite{humbatova20taxonomy} further refined the bug investigation and provided a taxonomy of real faults in deep learning systems.
The taxonomy was derived using 375 buggy program posts obtained from \sof and \gh designed using three popular  DL libraries: Tensorflow, Keras, and PyTorch.
Moreover, the taxonomy was further enhanced by conducting interviews with 20 researchers and validated by involving an additional 21 developers. 
The taxonomy is broadly classified into five categories: \textit{Model, GPU usage, API, Tensors \& Inputs, and Training}.
The structural bugs (bugs due to suboptimal model structure), might appear in any of the five categories and can cause crashes or poor/unexpected accuracy when the DL model is trained. 
The bugs in some of these categories, \ie \texttt{`GPU Usage'}, \texttt{`API'}, and \texttt{`Tensors \& Inputs'} cause the DL program to crash while bugs in categories \texttt{`Model'} and \texttt{`Training'} typically result in low/stuck accuracy during training.
While non-crashing bugs do not raise an exception, they negatively impact training and lead to poor generalization, crash bugs raise an exception during compilation/execution, \eg tensor shape mismatch, or deprecated API.
We relied on the taxonomy provided by \cite{humbatova20taxonomy} and focused on non-crashing bugs in this paper.

\begin{table}
\Large

\renewcommand{\arraystretch}{1.21}
\caption{Mapping between Different Types of Bugs, Dataset Characteristics, and Verification Rules. }
 \centering
 \begin{adjustbox}{width=1.0\textwidth}{

\begin{tabular}{|c|c|c|c|}
\hline
\textbf{\begin{tabular}[c]{@{}c@{}}Bug \\ Categories\end{tabular}}                            & \textbf{Type of Bug}                                                                                                  & \textbf{\begin{tabular}[c]{@{}c@{}}Dataset Characteristics \\ used to fix the bug \\ in SO posts\end{tabular}} & \textbf{Rules}                                                                                                                                                                                                             \\ \hline
\rowcolor[HTML]{D0D0D0} 
\begin{tabular}[c]{@{}c@{}}Activation \\ Function\end{tabular}                                     & \begin{tabular}[c]{@{}c@{}}Wrong type of activation/\\  Missing/redundant softmax \\  or relu activation\end{tabular} & \begin{tabular}[c]{@{}c@{}}Number of classes\\  Type of problem \\  (regression/classification)\end{tabular}   & \begin{tabular}[c]{@{}c@{}}Choice of Non-Linearity (CNL): \\  Checks for missing/redundant/wrong activation function\end{tabular}                                                                                          \\ \hline
                                                                                                   & Wrong filter for conv layer                                                                                           & \begin{tabular}[c]{@{}c@{}}Type of images \\ (RGB/Grayscale)\end{tabular}                                      & \begin{tabular}[c]{@{}c@{}}Inaccurate Number of Filters (INF): \\  Checks inappropriate number of filters for each conv layer\end{tabular}                                                                                 \\ \cline{2-4} 
                                                                                                   & \begin{tabular}[c]{@{}c@{}}Suboptimal number \\  of neurons\end{tabular}                                              & \begin{tabular}[c]{@{}c@{}}Type of images \\ (RGB/Grayscale)/\\  Type of problem\end{tabular}                  & \begin{tabular}[c]{@{}c@{}}Incorrect Number of Neurons (INN): \\  Detects incorrect number of units dense layers\end{tabular}                                                                                              \\ \cline{2-4} 
\multirow{-5}{*}{\begin{tabular}[c]{@{}c@{}}Layer \\ Properties\end{tabular}}                      & \begin{tabular}[c]{@{}c@{}}Wrong amount or \\  type of pooling\end{tabular}                                           & Any type of data/problem                                                                                       & \begin{tabular}[c]{@{}c@{}}Insufficient Downsampling (IDS):\\  Checks for inappropriate amount of pooling after conv layer\end{tabular}                                                                                    \\ \hline
\rowcolor[HTML]{D0D0D0} 
\cellcolor[HTML]{D0D0D0}                                                                           & Missing Dropout Layer                                                                                                 & Any type of data/problem                                                                                       & \begin{tabular}[c]{@{}c@{}}Missing or Redundant Dropout (MRD):\\  checks if dropout is applied after dense and conv layer\end{tabular}                                                                                     \\ \cline{2-4} 
\rowcolor[HTML]{D0D0D0} 
\multirow{-3}{*}{\cellcolor[HTML]{D0D0D0}\begin{tabular}[c]{@{}c@{}}Missing \\ Layer\end{tabular}} & \begin{tabular}[c]{@{}c@{}}Missing Normalization \\ Layer\end{tabular}                                                & Any type of data/problem                                                                                       & \begin{tabular}[c]{@{}c@{}}Missing Normalization Layer (MNL):\\  checks for missing normalization layer after dense and conv layers\end{tabular}                                                                           \\ \hline
\begin{tabular}[c]{@{}c@{}}Model \\ Properties\end{tabular}                                        & \begin{tabular}[c]{@{}c@{}}Suboptimal Network \\  Architecture\end{tabular}                                           & \begin{tabular}[c]{@{}c@{}}Type of images \\ (RGB/Grayscale)\end{tabular}                                      & \begin{tabular}[c]{@{}c@{}}Inappropriate Number of Convolution Layers (ICL): \\  checks for suboptimal conv layers\\  Improper Number of Fully Connected Layers (IFL):\\  checks for unnecessary dense layers\end{tabular} \\ \hline
\rowcolor[HTML]{D0D0D0} 
\begin{tabular}[c]{@{}c@{}}Preprocessing \\ of Data\end{tabular}                                   & Missing Preprocessing                                                                                                 & Any type of data/problem                                                                                       & \begin{tabular}[c]{@{}c@{}}Input Data not Normalized (IDN):\\  checks if the data is normalized or not\end{tabular}                                                                                                        \\ \hline
\begin{tabular}[c]{@{}c@{}}Loss \\ Function\end{tabular}                                           & \begin{tabular}[c]{@{}c@{}}Wrong selection of \\  loss function\end{tabular}                                          & \begin{tabular}[c]{@{}c@{}}Number of classes\\  Type of problem \\  (regression/classification)\end{tabular}   & \begin{tabular}[c]{@{}c@{}}Labels, output layer activation, and Loss Mismatch (LLM):\\  detects mismatch between output layer activation and loss function\end{tabular}                                                    \\ \hline
\rowcolor[HTML]{D0D0D0} 
\cellcolor[HTML]{D0D0D0}                                                                           & \begin{tabular}[c]{@{}c@{}}Suboptimal Learning \\ Rate\end{tabular}                                                   & Any type of data/problem                                                                                       & \begin{tabular}[c]{@{}c@{}}Learning Rate Out-of-Bound (LOB):\\  checks learning rate is in proper range\end{tabular}                                                                                                       \\ \cline{2-4} 
\rowcolor[HTML]{D0D0D0} 
\multirow{-3}{*}{\cellcolor[HTML]{D0D0D0}Hyperparameters}                                          & Suboptimal Batch Size                                                                                                 & Size of training set                                                                                           & \begin{tabular}[c]{@{}c@{}}Inadequate Batch Size (IBS):\\  checks for inadequate batch size\end{tabular}                                                                                                                   \\ \hline
\end{tabular}

}
\end{adjustbox}
\label{tab:mapping}
\end{table}

\subsection{Mapping between Dataset Characteristics and Structural Bugs}
\label{ssec:Mapping}
We manually inspected the dataset released by \cite{humbatova20taxonomy} and filtered out the posts related to non-crashing bugs.
We found 105 posts
in relation to our targeted bugs in DL programs. 
The DL programs derived from these posts include models with different architectures, such as FCNN and CNN, and different network depths. 
The distribution of DL models with varying depths obtained from 105 posts is shown in Fig.~\ref{ModelLayer}.
These posts are manually reviewed by authors to understand the debugging process followed by developers to identify the underlying cause of the bug, its symptoms, and the methods used to fix structural bugs.
We observe that most of the bugs related to the structure of the model, \eg wrong activation, and suboptimal neurons which do not cause the program to crash but lead to training issues, can be found at the beginning of the training process using the characteristics of the dataset, \eg the number of classes and/or type of problem, \ie regression or classification.
Therefore, we mapped
each structural bug with the dataset characteristics utilized to fix it.
Table~\ref{tab:mapping} shows the mapping between each type of bug and dataset characteristics used to fix the corresponding bug. 
Bugs and fixes are obtained using the DL models with different architectures, \ie FCNN and CNN designed for different tasks, such as image classification, text classification, multi-label classification, and regression, and with varying depths (Fig.~\ref{ModelLayer}),
which highlights the generalizability of using them for bug localization. 
Below, we discuss the manual labeling process in detail.

\paragraph{Manual Labeling} 
In the 105 posts obtained after inspecting the dataset of bugs released by Humbatova \etal~\cite{humbatova20taxonomy}, two authors independently reviewed these posts and classified the bugs into various categories following the taxonomy of \cite{humbatova20taxonomy}. Both authors identified seven bug categories: activation function, layer properties, missing layer, model properties, data preprocessing, loss functions, and hyperparameters. After discussion, we found that both authors reached 100\% agreement on categorizing these posts into respective categories. The next step is to understand the debugging process followed by the developers to identify the underlying cause of the bug, its symptoms, and the methods used to fix these bugs. During the distribution of the posts into different bug categories, two authors observed certain frequent terms such as type of data, type of task, number of classes, dimension of images, and size of the training dataset in these posts. These terms are used as initial labels to map the bugs and their fixes used in the posts. We followed the procedure described by Biswas \etal~\cite{biswas22art}, and two authors (raters) independently labeled these posts. After labeling all posts, we calculated the agreement using Cohen’s Kappa coefficient and conducted a discussion session between the raters and moderators (co-authors). We adopted Biswas \etal's~\cite{biswas22art} interpretation of Kappa ($\kappa = ([0, 1]$), the higher the better). After the first round, we found an almost perfect agreement ($\kappa = 0.85$). There were only a few disagreements about the labeling, which were resolved after a discussion session involving the raters and the moderators. After discussion, all the authors reached a perfect agreement ($\kappa = 1$). Finally, all the authors collectively examined each post for a final pass. The labels after the first round and final labels are provided in our repository \cite{myRepo}.

\subsection{Verification Rules Creation Methodology}
To reaffirm the bug's underlying source and its effect on performance, we reviewed the literature~\cite{LeCun98LeNet5,Krizhevsky12AlexNet,Simonyan14VGG,backprop92,Bengio12,Baker17,LeCun89,Krizhevsky2010,Krizhevsky09cifar10,Shea15}. 
These research papers provide several guidelines and design principles for designing DL programs. 
We used these guidelines and design principles to define the verification rules. Therefore, for each bug in Table~\ref{tab:mapping}, we define a verification rule (discussed in detail in Section~\ref{sec:Detection}). 
We also defined the thresholds for various rules using the fix suggestions obtained from 105 posts, illustrated in Section~\ref{ssec:Mapping}. 
Since the fix suggestions were effective for the DL models with different architectures, \ie FCNN and CNN designed for different tasks, such as image classification, text classification, multi-label classification, and regression, and with varying depths, we utilized them to define the thresholds.

We used the buggy DL programs 
provided in the defect4ML benchmark~\cite{Morovati23defects4ML} to verify these rules.
This benchmark has 100 faulty DL programs obtained from \sof and \gh belonging to different bug categories proposed in \cite{humbatova20taxonomy}.
We randomly picked one buggy DL program for each bug type supported by \ourapproach and verified the rules. As \ourapproach supports 12 types of bugs (shown in Table~\ref{tab:mapping}), we specifically selected 12 programs for this task.
This process helped us verify the correctness of defined rules and thresholds.

\section{Approach} \label{sec:Approach}

In this section, we first describe the analysis techniques used to detect the structural bugs.
Then, we discuss the verification rules and explain how these rules are used in 
our approach, \ourapproach, to automatically detect and localize structural bugs.

\begin{figure*}[ht!]
	\centering
	\includegraphics[width=1.0\linewidth,scale=2.0,trim={0cm 0.cm 0cm 0.0cm}]{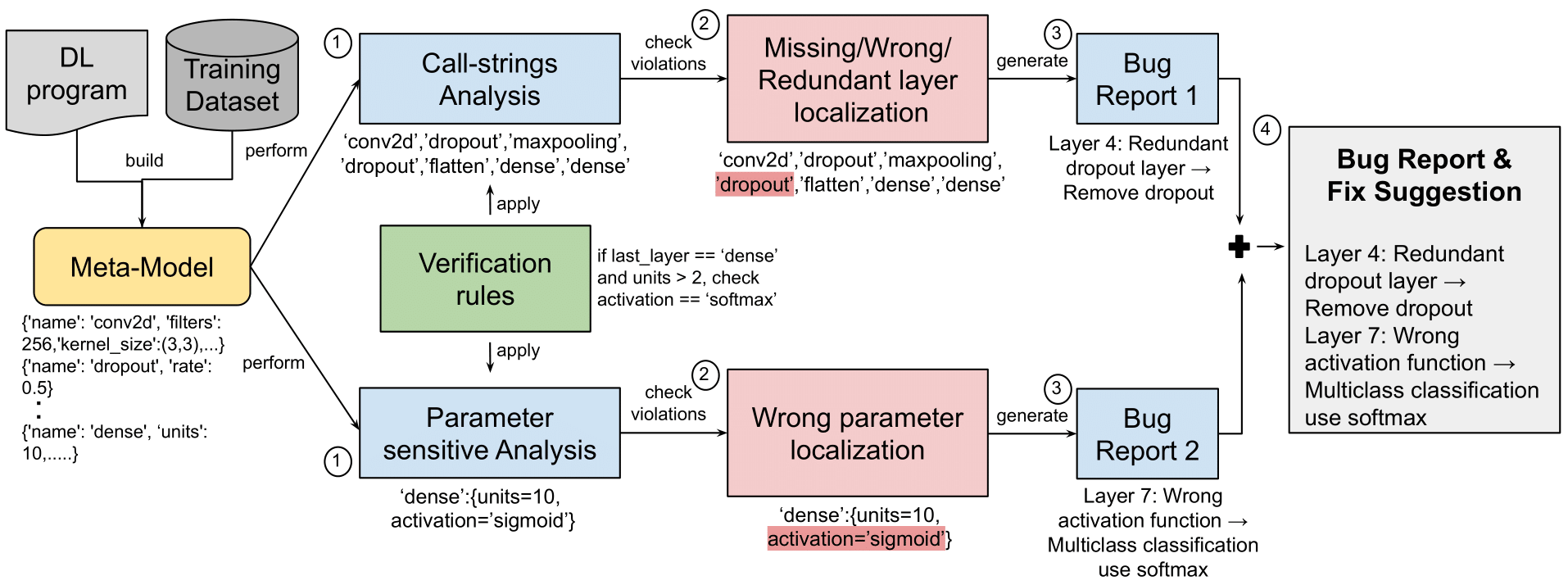}
    \caption{Overview of \ourapproach.}
	\label{Overview}
\end{figure*}

\subsection{Detecting Structural Bugs} \label{sec:Detection}
\begin{figure*}[ht!]
	\centering
	\includegraphics[width=1.0\linewidth,scale=2.0,trim={0cm 0.cm 0cm 0.0cm}]{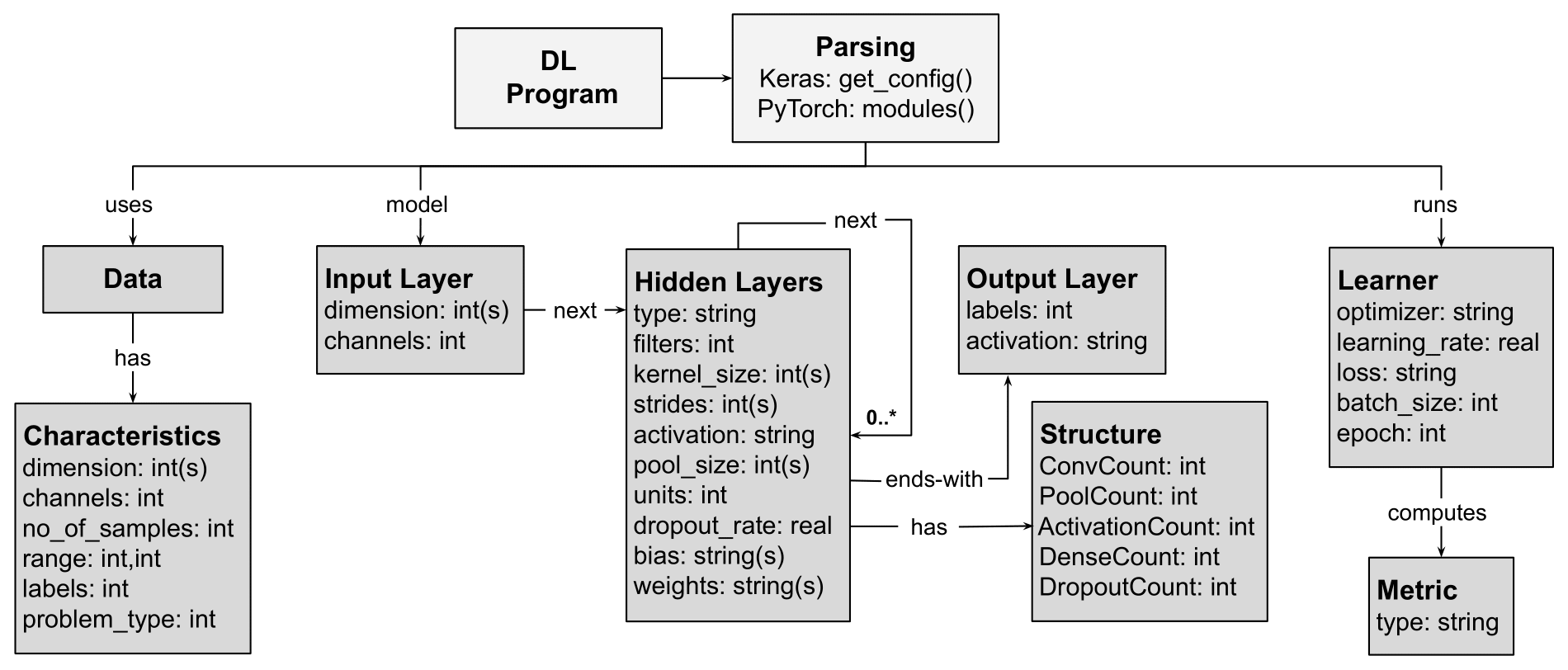}
    \caption{Meta-Model used in \ourapproach.}
	\label{meta-model}
\end{figure*}

Context-sensitive analysis is a common interprocedural analysis technique used to develop more efficient programs~\cite{Shivers91}.
Various context-sensitive analysis approaches have been proposed in the past, \eg call-strings approach~\cite{Sharir78}, functional approach~\cite{Sharir78}, call graphs~\cite{Grove97}.
Unlike traditional programs where one function can be called multiple times and one function can call another function,
DL programs follow different structures where the model is built sequentially by calling different layer APIs, \eg \texttt{Conv2D}, \texttt{Activation}, \texttt{MaxPooling2D}, provided by DL libraries one after the other.
For traditional programs, the call-strings approach is used to keep track of how many times each function is called, how many times it is returned, and what other functions are called by it.
In our approach, we used call strings to keep track of API calls and applied verification rules
to detect missing, redundant, or wrong API usage.
Parameter-sensitive approach~\cite{Wei15} is used for traditional programs to analyze each function call independently to determine how the function's parameters affect its functioning.
We utilized this approach to examine each API call and its parameters to identify incorrect parameters.
Fig.~\ref{Overview} shows an overview of \ourapproach.
We built upon the meta-model proposed by Nikanjam \etal~\cite{nikanjam2021neuralint} for DL programs, which captures the various components such as the architecture of the model, learner, and details related to shuffling and batching by parsing the model's source code. However, their meta-model does not capture the characteristics of the data.
In our approach, for an executable DL program, a meta-model is built as shown in Fig.~\ref{meta-model} which captures the characteristics of the dataset, \eg type of input, number of classes, and properties of the DL model, \eg filters in convolution layers, dropout rate.
We utilized the \texttt{get\_config()} and \texttt{modules()} APIs provided by \Keras and \PyTorch DL libraries, respectively, for parsing the configuration of the model.
On the meta-model, call-string analysis and parameter-sensitive analysis are performed and violations are checked using the verification rules discussed below. 
Violations are used to detect bugs and to keep track of layer numbers which are utilized in bug report for localization.
Each analysis generates a bug report that is combined to generate a final bug report with fix suggestions.
If \ourapproach detects a
bug, the training aborts with a report containing the bug's location and recommended fixes to alert the developer; otherwise, training continues. 
Below, we discuss the verification rule used by \ourapproach for detecting and localizing each bug shown in Table~\ref{tab:mapping}.

\subsubsection{Choice of Non-Linearity (CNL)}
\textbf{Rationale:} Convolution and Dense are linear operations; therefore, 
incorporating non-linear activation functions is crucial in the DL models to satisfy the Universal Approximation Theorem (UAT). Non-linearity is added to the output of convolution and dense layers of DL programs \via non-linear activation functions, which ensures these models learn from complex data patterns.
Activation functions can be saturating (\eg sigmoid and tanh) or non-saturating (\eg ReLU and its variants) \cite{ReLU}.
These activations transform the value of convolution and dense operation into a restricted range~\cite{Nwankpa18}; therefore, applying multiple or redundant activations will result in the wrong output in the last layer. Hence, for each convolution and dense layer activation function is used once.
Also, the choice of activation depends on the type of the task, \ie regression or classification.
For example, for image data,
Krizhevsky~\etal~\cite{Krizhevsky12AlexNet} have shown the benefit of using non-saturating non-linear activation functions for hidden layers over saturating counterparts.
First, non-saturating functions like ReLU help the network to learn faster, thus accelerating the training process.
Second, these activation functions help mitigate common training problems, such as exploding and vanishing gradients~\cite{wardat22DeepDiagnosis,Zhang21Autotrainer}.
Therefore, choosing the right activation function for hidden layers is crucial for enhancing the model's performance.

\textbf{Detection:} If the activation function for hidden layers, \ie convolution or dense layer is missing or multiple activation functions are used for the same layer, \ourapproach identifies it as a bug.  \ourapproach also considers the type of task for which the model is designed, checks incorrect usage of the activation function, and reports it as a bug. 

\subsubsection{Inaccurate Number of Filters (INF)}

\textbf{Rationale:} In CNN programs, features are learned by the convolution layers \via various filters in each layer.
Since the input has multiple features, distinct types of features are learned by these layers, starting from basic features to higher-order features as we go deeper into the network~\cite{LeCun98LeNet5}.
For example, for image data, the first few layers learn the basic features, \ie lines, edges, and corners and the deeper layers learn the higher-order features like objects.
The filters in each convolution layer depend on the type of input the CNNs are designed for.
For instance,
if the CNN model is designed for image classification, then the input 
images can be gray-scale or color.
Krizhevsky~\etal~\cite{Krizhevsky12AlexNet} and Simonyan~\etal~\cite{Simonyan14VGG} have shown the benefit of using more filters for color images 
as compared to the model designed for gray-scale images~\cite{LeCun98LeNet5} as color images have more complex features as compared to the gray-scale images. 
Convolution layers with too many filters often lead to overfitting of the model on training data, which restricts the model's ability to generalize adequately on test data.
While few filters impede the model's capacity to learn, which leads to poor performance during training and testing.
While designing CNN programs, 
the filters in the convolution layer must be configured by the developer by taking into account the training dataset characteristics, \ie type of data.

\textbf{Detection:} For detecting this bug, \ourapproach checks parameter \textit{filters} in \texttt{Conv1D} and \texttt{Conv2D} API and considers the dataset characteristic - \textit{channels} captured in meta-model. It checks if the \textit{filters} are less than 16 or more than 512 in each convolution layer for \textit{channels} = 3 (represents color images) or \textit{filters} are less than 6 or more than 256 in each convolution layer for \textit{channels} = 1 (represents gray-scale images or tabular data), the bug is reported with the fix location.

\subsubsection{Incorrect Number of Neurons (INN)}

\textbf{Rationale:} 
The width of the network, \ie number of neurons in each dense layer of the DL model is defined by considering the task for which the model is built. For instance, for classification tasks, the number of neurons depends on the number of classes for classification~\cite{Larochelle09}.
As these neurons learn features during training, a large number of neurons results in more trainable parameters.
Therefore, choosing the correct configuration helps in improving performance and results in faster training.
Also, for CNNs, the number of neurons in each layer should either remain the same or decrease while moving deeper 
toward the output layer \cite{LeCun98LeNet5,Krizhevsky12AlexNet}.

\textbf{Detection:} This bug is detected by \ourapproach by checking the \textit{units} in the dense layer excluding the output layer. The \textit{units} must be less than or equal to the size of the input each dense layer receives; otherwise \ourapproach reports it as a bug. For CNN programs, \ourapproach checks if the units in dense layers decreases progressively towards the output layer, reporting a bug if this condition is not met.

\subsubsection{Insufficient Downsampling (IDS)}

\textbf{Rationale:} In CNN programs, different filters are used by the convolution layer to generate feature maps \cite{LeCun98LeNet5}.
Feature maps extract the position of the features in the input and summarize the presence of features.
To make feature maps more robust and make them invariant to the local translation, downsampling is used \cite{LeCun98LeNet5}.
Downsampling helps in reducing the size of the feature maps while preserving large or important structural elements. 
Pooling is the commonly used method for downsampling \cite{Krizhevsky12AlexNet} and it is used after convolution layer(s) to make the model more robust against shifts and distortion~\cite{LeCun98LeNet5}.
Stacking several convolution layers without using pooling in between makes the model less robust to the local translation and affects the performance of the model.
Therefore, to make the model robust, the pooling is recommended to be applied after a stack of few convolution layers~\cite{Krizhevsky12AlexNet,Simonyan14VGG}. 

\textbf{Detection:} \ourapproach detects this bug by checking if the pooling layer is missing after 4 consecutive convolution layers.

\subsubsection{Missing or Redundant Dropout (MRD)}

\textbf{Rationale:} Dropout is a regularization technique used in DL programs to prevent the model from overfitting and thus helps in better generalization.
Srivastava~\etal~\cite{Srivastava14dropout} proposed this approach and has shown the effectiveness of using dropout on the performance of the DL model. 
\cite{Srivastava14dropout} suggests applying dropout after dense and convolution layers once as these layers have learnable parameters.

\textbf{Detection:} To detect this bug, \ourapproach checks if dropout is applied after dense and convolution layers. \ourapproach also counts the number of times dropout (DropoutCount) is applied to each dense and convolution layer. If DropoutCount for each layer is greater than 1, \ourapproach reports it as a bug.

\subsubsection{Missing Normalization Layer (MNL)}

\textbf{Rationale:} Batch Normalization is a technique used to train DL model faster. 
The goal of Batch Normalization
is to generate a consistent distribution of activation values throughout the training which helps in faster convergence.
Therefore, to train DL models faster, Batch Normalization is recommended after convolution and dense layers before applying non-linearity~\cite{Ioffe15BatchNormalization}.

\textbf{Detection:} \ourapproach detects this bug by keeping track of layers after dense and convolution layers. If the Batch Normalization layer is missing after these layers and before the activation layer, \ourapproach reports it as a bug.

\subsubsection{Inappropriate Number of Convolution Layers (ICL)}

\textbf{Rationale:} In CNN program, the convolution layers are used to extract the local features from the input.
The elementary visual features such as edges, lines, \etc are learned by the first few convolution layers~\cite{LeCun98LeNet5}.
And, subsequent convolution layers are used to learn the higher-order features by combining the features from the previous convolution layers~\cite{LeCun98LeNet5}.
State-of-the-art CNN architectures~\cite{Krizhevsky12AlexNet,Simonyan14VGG,He16ResNet} showed the advantage of having more convolution layers in the CNN model designed for training datasets with color images.
For instance, for color images in ImageNet~\cite{imagenet} dataset, popular CNN architectures, \eg AlexNet~\cite{Krizhevsky12AlexNet}, VGG~\cite{Simonyan14VGG} used more convolution layers to learn the features in contrast to fewer convolution layers used by LeNet-5~\cite{LeCun98LeNet5} for grayscale images in MNIST~\cite{LeCun98LeNet5} dataset.
Therefore, the number of convolution layers must be selected by considering the type of images in the training dataset, \ie grayscale or color.

\textbf{Detection:} This bug is detected by \ourapproach by counting the number of convolution layers. For a dataset with grayscale images, there must be at least 2 or more convolution layers, and for color images, there must be at least 3 or more convolution layers.

\subsubsection{Improper Number of Fully Connected Layers (IFL)}

\textbf{Rationale:} In CNN programs,
the fully connected layers are used for classification. 
These layers have a large number of trainable parameters, so more time and memory are required to train them.
It is advised~\cite{LeCun98LeNet5,Krizhevsky12AlexNet,Simonyan14VGG} to use one or two fully connected layers since they save training time, prevent the model from over-fitting, and improve generalization.

\textbf{Detection:} \ourapproach checks the number of dense layers used in the structure of the CNN model. If the number of dense layers is more than 3, \ourapproach warns the developer to reduce the number of fully connected layers.

\subsubsection{Input Data not Normalized (IDN)}
\textbf{Rationale:} Backpropagation is a popular algorithm used to train neural networks \cite{backprop92}.
The efficiency of the algorithm depends on the input data \cite{LeCun89}.
LeCun~\etal~\cite{LeCun12} provides several guidelines for more efficient back-propagation.
Normalization of the input is one of them.
If the input data to the model are close to zero, it results in faster convergence and thus, makes the training faster.
Therefore, in the data preprocessing stage, the training data must be normalized in order to achieve better performance.

\textbf{Detection:} \ourapproach detects this bug by checking the range of input values. If the \textit{range} is not between [0,1] or [-1,1], \ourapproach and alerts the developer about it by providing a message in the bug report.

\subsubsection{Labels, output layer activation, and Loss Mismatch (LLM)}

\textbf{Rationale:} For image classification, the activation function for the output layer is chosen based on the type of classification, \ie binary or multi-class classification.
And, for regression tasks, for the output layer, linear activation is preferred over non-linear activation~\cite{Larochelle09}. 
\textit{Loss} is used to evaluate the performance of the model and to compute the error at the time of training.
Cross-entropy is the commonly used loss function for classification problems.
As suggested by~\cite{Kerasloss, Pytorch}, for binary classification, it is preferable to use the sigmoid activation function in the output layer and binary cross-entropy as a loss function to compute the error.
For multi-class classification, it is suggested to use the softmax activation function in the output layer.
If the loss function is not selected according to the last layer activation function, then due to improper gradient, the model will learn inefficiently.

\textbf{Detection:} \ourapproach uses the problem type passed as input to the callbacks and checks the activation function and loss. If there is any mismatch as explained above, the bug is reported.

\begin{table}[]
\renewcommand{\arraystretch}{1.2}

\caption{Summary of Rules used in \ourapproach.}
 \centering
\begin{adjustbox}{width=1.0\textwidth}


\begin{tabular}{|c|c|c|c|l|}
\hline
\textbf{\begin{tabular}[c]{@{}c@{}}Bug \\ Type\end{tabular}} & \textbf{\begin{tabular}[c]{@{}c@{}}Model \\ Type\end{tabular}} & \textbf{\begin{tabular}[c]{@{}c@{}}Analysis \\ Technique\end{tabular}} & \textbf{\begin{tabular}[c]{@{}c@{}}API used for \\ Bug Detection\end{tabular}} & \multicolumn{1}{c|}{\textbf{Bug Detection Rule}}                                                                                                                                                                                                                                                                                                                                                                 \\ \hline
\rowcolor[HTML]{D0D0D0} 
CNL                                                          & \begin{tabular}[c]{@{}c@{}}FCNN/\\  CNN\end{tabular}           & call-strings                                                           & \begin{tabular}[c]{@{}c@{}}Dense(),\\  Conv1d(),\\  Conv2d()\end{tabular}      & \begin{tabular}[c]{@{}l@{}}if layer = `dense' or `conv' and activation\_count = 0 or \textgreater{}1\\ before next `dense' or `pooling' or `conv' layer\end{tabular}                                                                                                                                                                                                                                             \\ \hline
INF                                                          & CNN                                                            & parameter-sensitive                                                    & \begin{tabular}[c]{@{}c@{}}Conv1d(),\\  Conv2d()\end{tabular}                  & \begin{tabular}[c]{@{}l@{}}if input\_type = `color\_images', conv\_filters \textless 16 and conv\_filters \textgreater 512 or\\ if input\_type = `grayscale\_images' or `tabular', conv\_filters \textless 6 and conv\_filters \textgreater 256\end{tabular}                                                                                                                                                     \\ \hline
\rowcolor[HTML]{D0D0D0} 
INN                                                          & \begin{tabular}[c]{@{}c@{}}FCNN/\\  CNN\end{tabular}           & parameter-sensitive                                                    & Dense()                                                                        & if dense\_layer\_units \textgreater size of input of each layer                                                                                                                                                                                                                                                                                                                                                  \\ \hline
IDS                                                          & CNN                                                            & call-strings                                                           & \begin{tabular}[c]{@{}c@{}}Conv1d(),\\  Conv2d()\end{tabular}                  & if consecutive\_conv\_layer\_count \textgreater 4 and layer\_next != `pooling'                                                                                                                                                                                                                                                                                                                                   \\ \hline
\rowcolor[HTML]{D0D0D0} 
MDR                                                          & \begin{tabular}[c]{@{}c@{}}FCNN/\\  CNN\end{tabular}           & call-strings                                                           & \begin{tabular}[c]{@{}c@{}}Dense(),\\  Conv1d(),\\  Conv2d()\end{tabular}      & \begin{tabular}[c]{@{}l@{}}if layer\_hidden = `activation' and layer\_next != `dropout' or \\ if layer = `dense' or `conv' and dropout\_count \textgreater 1 before the next `dense' or `pooling' or `conv' layer\end{tabular}                                                                                                                                                                                   \\ \hline
MNL                                                          & \begin{tabular}[c]{@{}c@{}}FCNN/\\  CNN\end{tabular}           & call-strings                                                           & \begin{tabular}[c]{@{}c@{}}Dense(),\\  Conv1d(),\\  Conv2d()\end{tabular}      & if layer = `dense' or `conv' and layer\_next != `batch\_normalization'                                                                                                                                                                                                                                                                                                                                           \\ \hline
\rowcolor[HTML]{D0D0D0} 
ICL                                                          & CNN                                                            & call-strings                                                           & Conv2d()                                                                       & \begin{tabular}[c]{@{}l@{}}if input\_type = `color\_images', conv\_layer\_count \textless 3 or\\  if input\_type = `grayscale\_images', conv\_layer\_count \textless 2\end{tabular}                                                                                                                                                                                                                              \\ \hline
IFL                                                          & CNN                                                            & call-strings                                                           & Dense()                                                                        & if dense\_layer\_count \textgreater 3                                                                                                                                                                                                                                                                                                                                                                            \\ \hline
\rowcolor[HTML]{D0D0D0} 
IDN                                                          & \begin{tabular}[c]{@{}c@{}}FCNN/\\  CNN\end{tabular}           & parameter-sensitive                                                    & fit()                                                                          & if input\_range != {[}0,1{]} or {[}-1,1{]}                                                                                                                                                                                                                                                                                                                                                                       \\ \hline
LLM                                                          & \begin{tabular}[c]{@{}c@{}}FCNN/\\  CNN\end{tabular}           & parameter-sensitive                                                    & \begin{tabular}[c]{@{}c@{}}Dense(),\\  compile()\end{tabular}                  & \begin{tabular}[c]{@{}l@{}}if problem\_type = `binary\_classification', \\ output\_layer\_activation != `sigmoid' and loss != `binary\_crossentropy' or\\ if problem\_type = `multiclass\_classification', \\ output\_layer\_activation != `softmax' and loss != `categorical\_crossentropy' or\\ if problem\_type = `regression', output\_layer\_activation != `linear' and loss != `mse' or `mae'\end{tabular} \\ \hline
\rowcolor[HTML]{D0D0D0} 
LOB                                                          & \begin{tabular}[c]{@{}c@{}}FCNN/\\  CNN\end{tabular}           & parameter-sensitive                                                    & compile()                                                                      & if learning\_rate \textgreater 0.01 and learning\_rate \textless 0.0001                                                                                                                                                                                                                                                                                                                                          \\ \hline
IBS                                                          & \begin{tabular}[c]{@{}c@{}}FCNN/\\  CNN\end{tabular}           & parameter-sensitive                                                    & fit()                                                                          & if batch\_size \textless 32 and batch\_size \textgreater 256                                                                                                                                                                                                                                                                                                                                                     \\ \hline
\end{tabular}

\end{adjustbox}

\label{tab:summary}
\end{table}
\subsubsection{Learning Rate Out-of-Bound (LOB)}

\textbf{Rationale:} Learning rate is an important parameter 
that controls how much the model weights are adjusted with respect to the loss during backpropagation \cite{LeCun12}.
Too low learning rate increases the training time as the update towards the minima is very small.
Sometimes, it is also possible that due to a small learning rate, the training gets stuck on a sub-optimal solution or never converges~\cite{Zhang21Autotrainer}.
And, if the learning rate is set too high, the weight updates will be large which may result in an oscillating loss at the time of training~\cite{Zhang21Autotrainer}.

\textbf{Detection:} \ourapproach looks for an inappropriate learning rate by using the following threshold. If \textit{learning\_rate} is greater than 0.01 or \textit{learning\_rate} is less than 0.0001, it will be detected as a bug by \ourapproach.

\subsubsection{Inadequate Batch Size (IBS)}
\textbf{Rationale:} Batch size is an important training-related hyperparameter 
whose value impacts the performance of the model \cite{Zhang21Autotrainer}.
With a smaller batch size, even without looking at the complete training data, the model starts to learn, leading to oscillating loss and it is uncertain that the model will converge to the global optima~\cite{Zhang21Autotrainer}.
A larger batch size, on the other hand, might make the model get trapped in the local minima, which leads to poor generalization and low accuracy~\cite{Keskar16,Zhang21Autotrainer}. 
LeCun~\etal~\cite{LeCun12} and Bengio~\etal~\cite{Bengio12} suggest using 32 as the initial batch size and doubling it until 256.

\textbf{Detection:} \ourapproach detects this bug by checking the batch\_size used for training. If the batch size is less than 32 or greater than 256, \ourapproach warns the developer to use the appropriate batch size.

\IncMargin{1.0em}
\begin{algorithm}

	\caption{\ourapproach Algorithm}
	\label{alg:algorithm1}
	\DontPrintSemicolon
	\SetKwData{Left}{left}
	\SetKwData{Up}{up}
	\SetKwFunction{Forward}{Activation}
	\SetKwInOut{Input}{Input}
	\SetKwInOut{Output}{Output}
     \Indentp{-1em}
    	\Input{M, problem\_type, input\_type}
	    \Output{Report with bug location and actionable fix}
	\Indp\Indpp
	\BlankLine
	$bug\_report1 \leftarrow []$\;
        $bug\_report2 \leftarrow []$\;
        $layer\_name \leftarrow []$\;
        $layer\_config \leftarrow []$\;
        \For{$L$ \KwIn $M.Layers$}
        { 
	    layer\_names.append($L.Name$)
            layer\_config.append($L.get\_config()$)}   
        ConvCount = layer\_names.count(``conv2d'')\;
        DenseCount = layer\_names.count(``dense'')\;
	bug\_report1.append(IDN(M.input))\;
	\For{$i$ \KwIn range(len(layer\_names))}{ 
	    \If{$layer\_name[i]$ == ``conv2d''}
	    {
	       
	        bug\_report2.append(INF($input\_type$, $layer\_config[i].filters$))\;
               
	   }
	   
	   \If{$layer\_name[i]$ == ``maxpooling2d'' or $layer\_name[i]$ ``averagepooling2d''}
	   {
	       
	        bug\_report1.append(IDS($ConvCount$))\;
	        
	        }
	\If{$layer\_name[i]$ == ``dense''}
	   {
	        
	        \eIf{$layer\_name[i]$ is the last layer of $M$}
	        {
	            bug\_report2.append(LLM( $problem\_type$, $layer\_config[i].activation$, $M.loss$, $M.labels$))\;
	       }
	        {
	       
	        bug\_report2.append(INN($layer\_config[i].units$))\;
	       
	       }
	      }  
        \If{$layer\_name[i]$ == ``conv1d'' or $layer\_name[i]$ == ``conv2d'' or $layer\_name[i]$ == ``dense''}
        {
            bug\_report2.append(CNL($layer\_config[i].activation$))\;
            bug\_report1.append(MRD())\;
            bug\_report1.append(MNL())\;
        }
	    }
	    bug\_report1.append(ICL($input\_type$, $ConvCount$))\;
            bug\_report1.append(IFL($input\_type$, $DenseCount$))\;
	    bug\_report2.append(IBS($M.batch\_size$))\;
	    bug\_report2.append(LOB($M.learning\_rate$))\;
        final\_report = bug\_report1 + bug\_report2\;
	\If {final\_report != null}
	{
	    Abort training\;
	    \Return final\_report\;
	    }
	    
	\BlankLine
	\Indm\Indmm
	\Indp\Indpp
\end{algorithm}

Algorithm~\ref{alg:algorithm1} gives more details about the steps depicted in Fig.~\ref{Overview}.
The algorithm first captures the configuration of each layer of the DL model.
Then, it iterates through all the layers of the input model and applies the rules, discussed in Section~\ref{sec:Detection}, 
according to the type of the layer considering the type of problem and input. 
For example, for a conv2d layer, the rule \texttt{INF()}, is applied considering the type of input.
For the pooling layer,\texttt{IDS()} is used.
As dense layers are used as hidden layers and output layer in the model, the algorithm invokes the rule \texttt{INN()} for hidden dense layers, and for the output layer, \texttt{LLM()} is applied according to the problem type.
Then, the rules \texttt{CNL()}, \texttt{MRD()}, \texttt{MNL()}, which are common for different layers, \ie conv1d, conv2d, and dense layers are applied.
After looping through all the layers of the model,
if the model is designed for an image classification task, then
the two rules \texttt{ICL()} and \texttt{IFL()} are applied.
Finally, the rules \texttt{IBS()} and \texttt{LOB()} common for any architecture (FCNN or CNN) are invoked.
Each rule localizes different types of bugs discussed in Section~\ref{sec:Detection} and records an error message in a list, \textit{bug\_report1} or \textit{bug\_report2}.
The two bug reports, \textit{bug\_report1} and \textit{bug\_report2} are finally concatenated.  
If the \textit{final\_report} list is not empty, the training process is aborted and the messages, which contain the bug's location and actionable fix, are provided to the user.
Otherwise, the algorithm terminates and training starts normally.

The bug detection rules and the analysis techniques used to identify each bug in \ourapproach are summarized in Table~\ref{tab:summary}.

\section{Evaluation}\label{sec:evaluation}

In this section, we discuss the experimental setting and report an empirical evaluation to demonstrate the effectiveness of \ourapproach.

\subsection{Research Questions}
In this paper, we answer the following research questions.

\begin{itemize}
    \item \textbf{RQ1 (Evaluation):} How effective is \ourapproach in localizing and providing the actionable fixes for bugs in DL programs compared to state-of-the-art? 
   
    \item \textbf{RQ2 (Ablation)}: To what extent does \ourapproach detect each category of bugs correctly?
 
    \item \textbf{RQ3 (Limitation):} In which cases does our technique fail to detect and localize the bugs? 
  
 \end{itemize}

\subsection{Experimental setup}

\subsubsection{Implementation}

We implemented \ourapproach on top of \textit{Keras} 2.3.0~\cite{Keras}, \textit{TensorFlow} 2.1.0~\cite{Tensorflow}, and \PyTorch 1.13.1~\cite{Pytorch}. The meta-model is built by parsing the DL program. The data characteristics are obtained using training data provided as input to the DL program.
The configuration of the layers and learner are obtained by using \textit{get\_config()} and \textit{modules()} APIs provided by \Keras and \PyTorch, respectively.
Algorithm~\ref{alg:algorithm1} is implemented as a Python class which can be imported with Keras/PyTorch program.
We conducted all the experiments on a computer with a 4.2 GHz Quad-Core Intel Core i7 processor and 32 GB 2400 MHz DDR4 GB of RAM running the 64-bit MacOS X version 10.15.7.

\subsubsection{Benchmark}
We collected buggy DL programs developed using \Keras and \PyTorch from \sof posts to construct our benchmark. We followed~\cite{islam19} and used the keywords ``\texttt{bug},'' ``\texttt{poor performance},'' ``\texttt{CNN},'' ``\texttt{low accuracy}'' to search for posts. 
We obtained 172 posts.
In some posts, we found that pre-trained models provided by deep learning libraries were used. 
We removed such posts and obtained 63 posts. 
In most of the posts, the complete code is not provided by the developer.
Since \ourapproach needs a complete DL program for evaluation,
we considered the posts with full code script.
Therefore, we ended up with 40 posts~\cite{SO1,SO2,SO3,SO4,SO5,SO6,SO7,SO8,SO9,SO10,SO11,SO12,SO13,SO14,SO15,SO16,SO17, SO18, SO19, SO20, SO21,SO22,SO23,SO24,SO25,SO26,SO27,SO28,SO29,SO30,SO31,SO32,SO33,SO34,SO35,SO36,SO37,SO38,SO39,SO40}.
Additionally, we examined the benchmark of NeuraLint~\cite{nikanjam2021neuralint}. 
As NeuraLint supports both crash and performance bugs, we filtered out the programs with performance bugs. We obtained 9 programs with bad performance. As Theia needs a DL program with the dataset and DL model for evaluation, we obtained 4 programs (SO\# 50079585, 34311586, 51749207, 58844149) with datasets from \nlint's benchmark.
These programs were already included in our benchmark.
We also examined the artifacts provided by the previous works~\cite{wardat21DeepLocalize,ghanbari2023deepmufl}.
Due to the overlapping of programs in these benchmarks and 40 programs in our benchmark,
we found that these programs were already included in our benchmark (40 posts) during our filtration process.
In total, we have 40 buggy DL programs from \sof in our benchmark. 
The bug dataset contains 22 multi-class classifier models (16 CNNs and 3 FCNNs designed for image data, and 3 FCNNs designed for structural data), 13 binary classifier models (8 CNNs designed for image data, 2 CNNs designed for structural data, and 3 FCNNs designed for structural data), 3 regression models (3 FCNNs designed for structural data) and 1 multi-label classifier (1 FCNNs designed for structural data).  
The 40 programs in our benchmark are “unseen”, \ie these programs are not considered while determining the thresholds (using 105 programs) in verification rules.

\begin{table}[h]
\centering
  \captionof{table}{Performance Comparison of Buggy and Repaired Models Designed for Multiclass Classification Task.}
\scriptsize
\begin{tabular}{|c|l|c|cc|ccc|}
\hline
\multirow{2}{*}{\textbf{SNo.}} & \multicolumn{1}{c|}{\multirow{2}{*}{\textbf{SO \#}}} & \multirow{2}{*}{\textbf{Problem Type}} & \multicolumn{2}{c|}{\textbf{\begin{tabular}[c]{@{}c@{}}Performance\\  of Buggy Model\end{tabular}}}       & \multicolumn{3}{c|}{\textbf{\begin{tabular}[c]{@{}c@{}}Performance \\  After Applying Patch from SO\end{tabular}}}                                                                                       \\ \cline{4-8} 
                               & \multicolumn{1}{c|}{}                                &                                        & \multicolumn{1}{c|}{\textbf{Loss}} & \textbf{\begin{tabular}[c]{@{}c@{}}Accuracy\\  (in \%)\end{tabular}} & \multicolumn{1}{c|}{\textbf{Loss}} & \multicolumn{1}{c|}{\textbf{\begin{tabular}[c]{@{}c@{}}Accuracy\\  (in \%)\end{tabular}}} & \textbf{\begin{tabular}[c]{@{}c@{}}Improvement \\ (in \%)\end{tabular}} \\ \hline
1                              & 64522751                                             & Multiclass                             & \multicolumn{1}{c|}{2.761}         & 82.000                                                               & \multicolumn{1}{c|}{0.298}         & \multicolumn{1}{c|}{89.250}                                                               & 7.250  $\uparrow$                                                                 \\ \hline
2                              & 50079585                                             & Multiclass                             & \multicolumn{1}{c|}{0.482}         & 76.220                                                               & \multicolumn{1}{c|}{0.511}         & \multicolumn{1}{c|}{79.100}                                                               & 2.880 $\uparrow$                                                                   \\ \hline
3                              & 47272383                                             & Multiclass                             & \multicolumn{1}{c|}{0.990}         & 64.610                                                               & \multicolumn{1}{c|}{0.480}         & \multicolumn{1}{c|}{80.500}                                                               & 15.890 $\uparrow$                                                                  \\ \hline
4                              & 51118032                                             & Multiclass                             & \multicolumn{1}{c|}{2.309}         & 9.900                                                                & \multicolumn{1}{c|}{0.599}         & \multicolumn{1}{c|}{79.580}                                                               & 69.680  $\uparrow$                                                                 \\ \hline
5                              & 37229086                                             & Multiclass                             & \multicolumn{1}{c|}{0.384}         & 86.250                                                               & \multicolumn{1}{c|}{0.314}         & \multicolumn{1}{c|}{88.530}                                                               & 2.280   $\uparrow$                                                                 \\ \hline
6                              & 48594888                                             & Multiclass                             & \multicolumn{1}{c|}{0.765}         & 73.480                                                               & \multicolumn{1}{c|}{0.424}         & \multicolumn{1}{c|}{85.810}                                                               & 12.330 $\uparrow$                                                                  \\ \hline
7                              & 59325381                                             & Multiclass                             & \multicolumn{1}{c|}{0.010}         & 9.870                                                                & \multicolumn{1}{c|}{0.050}         & \multicolumn{1}{c|}{98.830}                                                               & 88.960  $\uparrow$                                                                 \\ \hline
8                              & 64188884                                             & Multiclass                             & \multicolumn{1}{c|}{1.236}         & 48.740                                                               & \multicolumn{1}{c|}{0.899}         & \multicolumn{1}{c|}{65.760}                                                               & 17.020  $\uparrow$                                                                 \\ \hline
9                              & 70554413                                             & Multiclass                             & \multicolumn{1}{c|}{2.770}         & 6.670                                                                & \multicolumn{1}{c|}{1.072}         & \multicolumn{1}{c|}{76.500}                                                               & 69.830 $\uparrow$                                                                  \\ \hline
10                             & 65275387                                             & Multiclass                             & \multicolumn{1}{c|}{1.946}         & 14.500                                                               & \multicolumn{1}{c|}{0.413}         & \multicolumn{1}{c|}{84.500}                                                               & 70.000  $\uparrow$                                                                 \\ \hline
11                             & 54923573                                             & Multiclass                             & \multicolumn{1}{c|}{0.151}         & 93.840                                                               & \multicolumn{1}{c|}{0.160}         & \multicolumn{1}{c|}{94.000}                                                               & 0.160   $\uparrow$                                                                 \\ \hline
12                             & 63027146                                             & Multiclass                             & \multicolumn{1}{c|}{2.303}         & 9.890                                                                & \multicolumn{1}{c|}{0.253}         & \multicolumn{1}{c|}{91.000}                                                               & 81.110 $\uparrow$                                                                   \\ \hline
13                             & 65659888                                             & Multiclass                             & \multicolumn{1}{c|}{0.676}         & 49.900                                                               & \multicolumn{1}{c|}{0.076}         & \multicolumn{1}{c|}{90.100}                                                               & 40.200  $\uparrow$                                                                 \\ \hline
14                             & 58666904                                             & Multiclass                             & \multicolumn{2}{c|}{Crash}                                                                                & \multicolumn{1}{c|}{0.112}         & \multicolumn{1}{c|}{70.100}                                                               & 70.100  $\uparrow$                                                                 \\ \hline
15                             & 55198221                                             & Multiclass                             & \multicolumn{1}{c|}{0.998}         & 60.320                                                               & \multicolumn{1}{c|}{0.935}         & \multicolumn{1}{c|}{63.840}                                                               & 3.520  $\uparrow$                                                                  \\ \hline
16                             & 55343875                                             & Multiclass                             & \multicolumn{1}{c|}{0.064}         & 97.240                                                               & \multicolumn{1}{c|}{0.051}         & \multicolumn{1}{c|}{98.290}                                                               & 1.050  $\uparrow$                                                                  \\ \hline
17                             & 38648195                                             & Multiclass                             & \multicolumn{1}{c|}{0.201}         & 47.810                                                               & \multicolumn{1}{c|}{0.727}         & \multicolumn{1}{c|}{68.190}                                                               & 20.380  $\uparrow$                                                                 \\ \hline
18                             & \multicolumn{1}{c|}{48385830}                        & Multiclass                             & \multicolumn{1}{c|}{nan}           & 9.870                                                                & \multicolumn{1}{c|}{0.290}         & \multicolumn{1}{c|}{91.200}                                                               & 81.330   $\uparrow$                                                                \\ \hline
19                             & 51930566                                             & Multiclass                             & \multicolumn{1}{c|}{0.808}         & 63.330                                                               & \multicolumn{1}{c|}{0.454}         & \multicolumn{1}{c|}{91.670}                                                               & 28.340  $\uparrow$                                                                 \\ \hline
20                             & 55328966                                             & Multiclass                             & \multicolumn{1}{c|}{nan}           & 9.860                                                                & \multicolumn{1}{c|}{0.008}         & \multicolumn{1}{c|}{99.770}                                                               & 89.910  $\uparrow$                                                                 \\ \hline
21                             & 58609115                                             & Multiclass                             & \multicolumn{1}{c|}{0.016}         & 99.780                                                               & \multicolumn{1}{c|}{0.070}         & \multicolumn{1}{c|}{99.800}                                                               & 0.020   $\uparrow$                                                                 \\ \hline
22                             & 59278771                                             & Multiclass                             & \multicolumn{1}{c|}{0.107}         & 50.100                                                               & \multicolumn{1}{c|}{0.059}         & \multicolumn{1}{c|}{96.670}                                                               & 46.570   $\uparrow$                                                                \\ \hline

\end{tabular}

\begin{tablenotes}
 \centering
        \fontsize{7pt}{7pt}\selectfont
        \item improvement \% = accuracy (in \%) after fix - accuracy (in \%) of buggy model.
        \item $\uparrow$ represents increase percentage, $\downarrow$ represents decrease percentage.

	\end{tablenotes}
  \label{tab:patchMulticlass}
\end{table}
\subsubsection{Results Representation}
As discussed in Section~\ref{sec:intro}, among the existing approaches for localizing bugs in the DL programs, only \nlint detects bugs before training and also supports CNN architecture-related bugs.
Therefore, we compared \ourapproach with \nlint on 40 buggy DL programs in our benchmark.
Table~\ref{tab:multiclass} summarizes the results of evaluating \ourapproach and NeuraLint for a multiclass classification task.
And, Table~\ref{tab:binary} summarizes the results of evaluating \ourapproach and NeuraLint for a binary classification and regression task.
In both the tables, Table~\ref{tab:multiclass} and Table~\ref{tab:binary}, for each DL program, we categorized the buggy programs into different categories indicated by ``Bug type'' which is obtained by mapping to the bugs specified in our verification rules, 
``Problem Type'' provides the details about the type of task, \ie regression or classification,
``SO \#'' represents the \sof post \#, 
``Recommended Fix from SO'' describes the recommended fix provided by the other users on \sof. 
The next two columns represent the results of \ourapproach and \nlint on our benchmark.
The well-known practice is to perform an evaluation on \sof posts with accepted answers~\cite{nikanjam2021neuralint}. It guarantees the suggested fix is a real fix for the problem and can be used as ground truth for evaluation.
In our benchmark, we found 6 posts from \sof without accepted answers.
We observed that some of the answers in these posts are marked as useful by users.
We considered them as a fix for the problem mentioned in the post.
To verify that the recommended fix effectively addressed the issue outlined in the post, 
we evaluated the performance of the buggy model before and after applying the suggested patch/fix.
Loss and accuracy are the common metrics used to evaluate the performance of the DL models.
We manually fixed the model following the suggestions from the accepted/useful answers of posts and computed the loss and accuracy before and after applying the fix shown in Table~\ref{tab:patchMulticlass} and Table~\ref{tab:patchBinary}.
We found that the recommended fix aided in resolving the issue described in the post in 35 out of 40 posts in our benchmark. 
In 5 out of 40 posts, we observed that the accuracy did not improve much following the fix suggestions from the accepted/useful answers. Therefore, two authors further investigated these posts and found that some fix suggestions are not marked as accepted or useful by developers.
However, upon applying these fixes, the two authors found that these patches helped in improving the model's performance.
We considered these fix suggestions as correct fixes and included them in the ground truth. These patches led to an average performance improvement of 38\% across 40 buggy DL programs. 
In Table~\ref{tab:multiclass} and Table~\ref{tab:binary}, the column labeled ``Recommended Fix from SO'' serves as the ground truth, which is used to determine the number of true positive and false negative cases.
Both the approaches, \ourapproach, and \nlint, also detect the bugs that are not recommended by \sof users.
For analyzing these results, we adopted the approach used by Nikanjam \etal \cite{nikanjam2021neuralint}.
Two authors independently checked the output and examined the DL program. We found that some structural inefficiencies are not pointed as a fix by any \sof user but are trivial and result in abnormal behavior during training, \eg multiple activation functions or dropout used for convolution or dense layers, missing pooling layer. 
For instance, in SO\# 47272383, we observed that multiple dropouts are used for the same convolution layer, which is not reported as a fix by the \sof user. The removal of one dropout layer, combined with the suggested fix from the \sof user led to an improvement in the model's performance.
We do not consider such fixes as false positives as addressing these structural inefficiencies helps improve the DL program's structure as discussed in Section~\ref{sec:Detection} which in turn helps improve the model's performance.
Therefore, we have not encountered any false positive cases and do not report them in Table~\ref{tab:multiclass} and Table~\ref{tab:binary}.
For both the approaches \ourapproach and \nlint, the ``Yes'' indicates whether the bug is identified and localized successfully or not. 
``--'' denotes that the target problem is not yet supported by the approach.

\begin{table}[t!]
 \centering
   \captionof{table}{Performance Comparison of Buggy and Repaired Models Designed for Binary Classification \& Regression Task.}
\scriptsize
\begin{tabular}{|c|l|c|cc|ccc|}
\hline
\multirow{2}{*}{\textbf{SNo.}} & \multicolumn{1}{c|}{\multirow{2}{*}{\textbf{SO \#}}} & \multirow{2}{*}{\textbf{Problem Type}}                                & \multicolumn{2}{c|}{\textbf{\begin{tabular}[c]{@{}c@{}}Performance\\  of Buggy Model\end{tabular}}}       & \multicolumn{3}{c|}{\textbf{\begin{tabular}[c]{@{}c@{}}Performance \\  After Applying Patch from SO\end{tabular}}}                                                                                       \\ \cline{4-8} 
                               & \multicolumn{1}{c|}{}                                &                                                                       & \multicolumn{1}{c|}{\textbf{Loss}} & \textbf{\begin{tabular}[c]{@{}c@{}}Accuracy\\  (in \%)\end{tabular}} & \multicolumn{1}{c|}{\textbf{Loss}} & \multicolumn{1}{c|}{\textbf{\begin{tabular}[c]{@{}c@{}}Accuracy\\  (in \%)\end{tabular}}} & \textbf{\begin{tabular}[c]{@{}c@{}}Improvement \\ (in \%)\end{tabular}} \\ \hline
1                              & 58844149                                             & Binary                                                                & \multicolumn{1}{c|}{7.645}         & 49.860                                                               & \multicolumn{1}{c|}{0.485}         & \multicolumn{1}{c|}{76.030}                                                               & 26.170  $\uparrow$                                                                 \\ \hline
2                              & 60261103                                             & Binary                                                                & \multicolumn{1}{c|}{0.894}         & 50.700                                                               & \multicolumn{1}{c|}{0.100}         & \multicolumn{1}{c|}{96.400}                                                               & 45.700     $\uparrow$                                                              \\ \hline
3                              & 56914715                                             & Binary                                                                & \multicolumn{1}{c|}{7.682}         & 49.630                                                               & \multicolumn{1}{c|}{0.490}         & \multicolumn{1}{c|}{84.750}                                                               & 35.120 $\uparrow$                                                                  \\ \hline
4                              & 60003876                                             & Binary                                                                & \multicolumn{2}{c|}{Crash}                                                                                & \multicolumn{1}{c|}{0.938}         & \multicolumn{1}{c|}{50.000}                                                               & 50.000  $\uparrow$                                                                 \\ \hline
5                              & 70428592                                             & Binary                                                                & \multicolumn{2}{c|}{Crash}                                                                                & \multicolumn{1}{c|}{0.000}         & \multicolumn{1}{c|}{98.100}                                                               & 98.100 $\uparrow$                                                                  \\ \hline
6                              & 40045159                                             & Binary                                                                & \multicolumn{1}{c|}{0.490}         & 72.880                                                               & \multicolumn{1}{c|}{0.441}         & \multicolumn{1}{c|}{80.100}                                                               & 7.220   $\uparrow$                                                                 \\ \hline
7                              & 45378493                                             & Binary                                                                & \multicolumn{1}{c|}{7.620}         & 50.000                                                               & \multicolumn{1}{c|}{0.076}         & \multicolumn{1}{c|}{99.000}                                                               & 49.000 $\uparrow$                                                                  \\ \hline
8                              & 51749207                                             & Binary                                                                & \multicolumn{1}{c|}{7.655}         & 49.800                                                               & \multicolumn{1}{c|}{0.011}         & \multicolumn{1}{c|}{99.600}                                                               & 49.800  $\uparrow$                                                                 \\ \hline
9                              & 58844149                                             & Binary                                                                & \multicolumn{1}{c|}{7.645}         & 49.860                                                               & \multicolumn{1}{c|}{0.485}         & \multicolumn{1}{c|}{76.030}                                                               & 26.170  $\uparrow$                                                                 \\ \hline
10                             & 31880720                                             & Binary                                                                & \multicolumn{1}{c|}{7.660}         & 50.000                                                               & \multicolumn{1}{c|}{0.005}         & \multicolumn{1}{c|}{99.900}                                                               & 49.900  $\uparrow$                                                                 \\ \hline
11                             & 39525358                                             & Binary                                                                & \multicolumn{1}{c|}{0.670}         & 61.590                                                               & \multicolumn{1}{c|}{0.575}         & \multicolumn{1}{c|}{92.310}                                                               & 30.720  $\uparrow$                                                                 \\ \hline
12                             & 31627380                                             & Binary                                                                & \multicolumn{1}{c|}{9.797}         & 39.040                                                               & \multicolumn{1}{c|}{0.643}         & \multicolumn{1}{c|}{68.120}                                                               & 29.080 $\uparrow$                                                                  \\ \hline
13                             & 34673164                                             & Binary                                                                & \multicolumn{1}{c|}{0.128}         & 77.780                                                               & \multicolumn{1}{c|}{0.422}         & \multicolumn{1}{c|}{88.890}                                                               & 11.110  $\uparrow$                                                                 \\ \hline
14                             & 34311586                                             & Regression                                                            & \multicolumn{1}{c|}{0.667}         & 33.300                                                               & \multicolumn{1}{c|}{0.684}         & \multicolumn{1}{c|}{66.670}                                                               & 33.370   $\uparrow$                                                                \\ \hline
15                             & 48221692                                             & Regression                                                            & \multicolumn{1}{c|}{2288.030}      & --                                                                   & \multicolumn{1}{c|}{95.283}        & \multicolumn{1}{c|}{--}                                                                   & 2192.747  $\downarrow$                                                              \\ \hline
16                             & 48251943                                             & Regression                                                            & \multicolumn{1}{c|}{736.928}       & --                                                                   & \multicolumn{1}{c|}{1.84 $\times 10^{-5}$}   & \multicolumn{1}{c|}{--}                                                                   & 736.928  $\downarrow$                                                               \\ \hline
17                             & 48934338                                             & Regression                                                            & \multicolumn{1}{c|}{1354.247}      & --                                                                   & \multicolumn{1}{c|}{248.703}       & \multicolumn{1}{c|}{--}                                                                   & 1105.544   $\downarrow$                                                              \\ \hline
18                             & \multicolumn{1}{c|}{44164749}                        & \begin{tabular}[c]{@{}c@{}}Multi Label\\  classification\end{tabular} & \multicolumn{1}{c|}{nan}           & 29.630                                                               & \multicolumn{1}{c|}{0.449}         & \multicolumn{1}{c|}{79.210}                                                               & 49.580  $\uparrow$                                                                 \\ \hline

\end{tabular}

\begin{tablenotes}
 \centering
        \fontsize{7pt}{7pt}\selectfont
        \item improvement \% = accuracy (in \%) after fix - accuracy (in \%) of buggy model (for classification).
        \item improvement  = loss after fix - loss of buggy model (for regression).
        \item $\uparrow$ represents increase percentage, $\downarrow$ represents decrease percentage.

	\end{tablenotes}
  \label{tab:patchBinary}
\end{table}
\begin{table}[hbt!]
\renewcommand{\arraystretch}{1.1}
\caption{Comparison of Bugs Localized by Theia and NeuraLint in Buggy DL Programs Designed for Multiclass Classification Task.}
 \centering
\begin{adjustbox}{width=1.0\textwidth}

\begin{tabular}{|ccccl|ccc|ccc|}
\hline
\multicolumn{1}{|c|}{}                                             & \multicolumn{1}{c|}{}                                    & \multicolumn{1}{c|}{}                                                     & \multicolumn{1}{c|}{}                                                   & \multicolumn{1}{c|}{}                                                   & \multicolumn{3}{c|}{\textbf{Theia}}                                                                                                                               & \multicolumn{3}{c|}{\textbf{NeuraLint}}                                                                                                                           \\ \cline{6-11} 
\multicolumn{1}{|c|}{\multirow{-2}{*}{\textbf{SNo.}}}              & \multicolumn{1}{c|}{\multirow{-2}{*}{\textbf{Bug Type}}} & \multicolumn{1}{c|}{\multirow{-2}{*}{\textbf{Problem Type}}}              & \multicolumn{1}{c|}{\multirow{-2}{*}{\textbf{SO \#}}}                   & \multicolumn{1}{c|}{\multirow{-2}{*}{\textbf{Recommended Fix from SO}}} & \multicolumn{1}{c|}{\textbf{Identify Bug}}       & \multicolumn{1}{c|}{\textbf{TP}}                                 & \textbf{FN}                                 & \multicolumn{1}{c|}{\textbf{Identify Bug}}       & \multicolumn{1}{c|}{\textbf{TP}}                                 & \textbf{FN}                                 \\
\hline
\end{tabular}

\end{adjustbox}

\label{tab:multiclass}
\end{table}

\begin{table}[h]
\renewcommand{\arraystretch}{1.1}
\caption{Comparison of Bugs Localized by Theia and NeuraLint in Buggy DL Programs Designed for Binary Classification \& Regression Task.}
 \centering
\begin{adjustbox}{width=1.0\textwidth}

%
}} & \multicolumn{1}{c|}{\multirow{-2}{*}{44164749}}                         & Change loss function                                                    & \multicolumn{1}{c|}{No}                          & \multicolumn{1}{c|}{\multirow{-2}{*}{0}}                         & \multirow{-2}{*}{2}                         & \multicolumn{1}{c|}{No}                          & \multicolumn{1}{c|}{\multirow{-2}{*}{0}}                          & \multirow{-2}{*}{2}                          \\ \hline
\multicolumn{5}{|c|}{\textbf{Total}}                                                                                                                                                                                                                                                                                                                                                             & \multicolumn{1}{c|}{\textbf{}}                   & \multicolumn{1}{c|}{\textbf{23}}                                 & \textbf{7}                                  & \multicolumn{1}{c|}{\textbf{}}                   & \multicolumn{1}{c|}{\textbf{4}}                                   & \textbf{22}                                  \\ \hline
\end{tabular}

\end{adjustbox}

\label{tab:binary}
\end{table}
\subsection{RQ1 (Evaluation)}
\subsubsection{Evaluation on Multiclass Classification Task}

We evaluated our approach and compared the state-of-the-art approach, \ie NeuraLint~\cite{nikanjam2021neuralint}, and reported the results in Table~\ref{tab:multiclass}. 
Below, we discuss the different categories of bugs and how NeuraLint performs compared to our approach.
For bug type \textit{LLM}, there are 14 programs in Table~\ref{tab:multiclass}. 
\ourapproach identified this bug in 14/14 programs (12 Keras and 2 PyTorch programs). To detect this bug, \ourapproach considers the type of problem, \eg binary classification, and multiclass classification, the last layer activation, and the loss function for which the DL model is built. 
Whereas,
\nlint supports this bug type and checks whether the loss function is correctly defined considering the last layer activation function. 
As discussed in Section~\ref{sec:Detection}, the last layer activation function and loss functions are defined according to the type of problem, \eg binary classification, and multiclass classification. As \nlint does not consider the type of problem while detecting this bug, it failed to detect bugs in SO \# \texttt{65275387, 54923573, 55198221, 48385830, 51930566, 55328966, 59278771}. The bugs in SO \# \texttt{65659888, 58666904} belong to \textit{LLM}, as these are PyTorch programs, \nlint does not support PyTorch programs.
For programs SO \# \texttt{37229086, 63027146, 48385830}, our results (in Table~\ref{tab:multiclass}) show that by considering the characteristics of the dataset, \ourapproach is able to detect all the bugs belonging to different categories \textit{ICL, IDN, INF, LLM, CNL}, whereas, \nlint failed to detect all the bugs and detected 2/7 bugs in 3 programs.
In total, \ourapproach detected 34/45 bugs found in 22 buggy real-world programs, whereas, \nlint detected 13/45 bugs.

\begin{table}[h]
 \centering
   \captionof{table}{Comparison of Impact of Actionable Fixes by Theia and NeuraLint on Buggy Models Performance Designed for Multiclass Classification Task.}
   
  \scalebox{0.74}{
\begin{tabular}{|c|c|cc|cccccccc|}
\hline
                                &                                  & \multicolumn{2}{c|}{}                                                                                                     & \multicolumn{8}{c|}{\textbf{\begin{tabular}[c]{@{}c@{}}Performance\\  After Fix\end{tabular}}}                                                                                                                                                                                                                                                                                                                                                                                                                                                                  \\ \cline{5-12} 
                                &                                  & \multicolumn{2}{c|}{\multirow{-2}{*}{\textbf{\begin{tabular}[c]{@{}c@{}}Performance\\  of Buggy Model\end{tabular}}}}     & \multicolumn{4}{c|}{\textbf{NeuraLint}}                                                                                                                                                                                                                                                           & \multicolumn{4}{c|}{\textbf{Theia}}                                                                                                                                                                                                                         \\ \cline{3-12} 
\multirow{-3}{*}{\textbf{SNo.}} & \multirow{-3}{*}{\textbf{SO \#}} & \multicolumn{1}{c|}{\textbf{Loss}}                 & \textbf{\begin{tabular}[c]{@{}c@{}}Accuracy\\  (in \%)\end{tabular}} & \multicolumn{1}{c|}{\textbf{Loss}}                 & \multicolumn{1}{c|}{\textbf{\begin{tabular}[c]{@{}c@{}}Accuracy\\  (in \%)\end{tabular}}} & \multicolumn{1}{c|}{\textbf{\begin{tabular}[c]{@{}c@{}}Improvement\\   (in \%)\end{tabular}}} & \multicolumn{1}{c|}{\textbf{FP}}                 & \multicolumn{1}{c|}{\textbf{Loss}}                 & \multicolumn{1}{c|}{\textbf{\begin{tabular}[c]{@{}c@{}}Accuracy\\  (in \%)\end{tabular}}} & \multicolumn{1}{c|}{\textbf{\begin{tabular}[c]{@{}c@{}}Improvement\\  (in \%)\end{tabular}}} & \textbf{FP} \\ \hline
1                               & 64522751                         & \multicolumn{1}{c|}{2.761}                         & 82.000                                                               & \multicolumn{1}{c|}{2.327}                         & \multicolumn{1}{c|}{9.780}                                                                & \multicolumn{1}{c|}{-72.220 $\downarrow$}                                                                  & \multicolumn{1}{c|}{Yes}                         & \multicolumn{1}{c|}{0.510}                         & \multicolumn{1}{c|}{82.190}                                                               & \multicolumn{1}{c|}{0.190 $\uparrow$}                                                                   & No          \\ \hline
2                               & 50079585                         & \multicolumn{1}{c|}{0.482}                         & 76.220                                                               & \multicolumn{1}{c|}{0.841}                         & \multicolumn{1}{c|}{64.380}                                                               & \multicolumn{1}{c|}{-11.840 $\downarrow$}                                                                  & \multicolumn{1}{c|}{Yes}                         & \multicolumn{1}{c|}{0.508}                         & \multicolumn{1}{c|}{80.890}                                                               & \multicolumn{1}{c|}{4.670 $\uparrow$}                                                                   & No          \\ \hline
3                               & 47272383                         & \multicolumn{1}{c|}{0.990}                         & 64.610                                                               & \multicolumn{1}{c|}{1.776}                         & \multicolumn{1}{c|}{19.810}                                                               & \multicolumn{1}{c|}{-44.800 $\downarrow$}                                                                  & \multicolumn{1}{c|}{Yes}                         & \multicolumn{1}{c|}{0.380}                         & \multicolumn{1}{c|}{88.420}                                                               & \multicolumn{1}{c|}{23.810 $\uparrow$}                                                                  & No          \\ \hline
4                               & 51118032                         & \multicolumn{1}{c|}{2.309}                         & 9.900                                                                & \multicolumn{1}{c|}{2.309}                         & \multicolumn{1}{c|}{9.900}                                                                & \multicolumn{1}{c|}{0.000 $\rightarrow$}                                                                    & \multicolumn{1}{c|}{No}                          & \multicolumn{1}{c|}{0.509}                         & \multicolumn{1}{c|}{82.020}                                                               & \multicolumn{1}{c|}{72.120 $\uparrow$}                                                                  & No          \\ \hline
\rowcolor[HTML]{D9D9D9} 
5                               & 37229086                         & \multicolumn{1}{c|}{\cellcolor[HTML]{D9D9D9}0.384} & 86.250                                                               & \multicolumn{1}{c|}{\cellcolor[HTML]{D9D9D9}0.107} & \multicolumn{1}{c|}{\cellcolor[HTML]{D9D9D9}96.370}                                       & \multicolumn{1}{c|}{\cellcolor[HTML]{D9D9D9}10.120 $\uparrow$}                                           & \multicolumn{1}{c|}{\cellcolor[HTML]{D9D9D9}No}  & \multicolumn{1}{c|}{\cellcolor[HTML]{D9D9D9}0.616} & \multicolumn{1}{c|}{\cellcolor[HTML]{D9D9D9}78.560}                                       & \multicolumn{1}{c|}{\cellcolor[HTML]{D9D9D9}-7.690 $\downarrow$}                                          & Yes         \\ \hline
6                               & 48594888                         & \multicolumn{1}{c|}{0.765}                         & 73.480                                                               & \multicolumn{1}{c|}{0.772}                         & \multicolumn{1}{c|}{73.160}                                                               & \multicolumn{1}{c|}{-0.320 $\downarrow$}                                                                   & \multicolumn{1}{c|}{Yes}                         & \multicolumn{1}{c|}{0.571}                         & \multicolumn{1}{c|}{79.920}                                                               & \multicolumn{1}{c|}{\cellcolor[HTML]{FFFFFF}6.440$\uparrow$}                                           & No          \\ \hline
7                               & 59325381                         & \multicolumn{1}{c|}{0.010}                         & 9.870                                                                & \multicolumn{1}{c|}{0.040}                         & \multicolumn{1}{c|}{9.870}                                                                & \multicolumn{1}{c|}{0.000 $\rightarrow$}                                                                    & \multicolumn{1}{c|}{No}                          & \multicolumn{1}{c|}{0.023}                         & \multicolumn{1}{c|}{99.340}                                                               & \multicolumn{1}{c|}{\cellcolor[HTML]{FFFFFF}89.470 $\uparrow$}                                          & No          \\ \hline
8                               & 64188884                         & \multicolumn{1}{c|}{1.236}                         & 48.740                                                               & \multicolumn{1}{c|}{1.111}                         & \multicolumn{1}{c|}{54.000}                                                               & \multicolumn{1}{c|}{5.260 $\uparrow$}                                                                    & \multicolumn{1}{c|}{No}                          & \multicolumn{1}{c|}{0.946}                         & \multicolumn{1}{c|}{62.590}                                                               & \multicolumn{1}{c|}{\cellcolor[HTML]{FFFFFF}13.850 $\uparrow$}                                          & No          \\ \hline
9                               & 70554413                         & \multicolumn{1}{c|}{2.770}                         & 6.670                                                                & \multicolumn{1}{c|}{2.770}                         & \multicolumn{1}{c|}{6.670}                                                                & \multicolumn{1}{c|}{0.000 $\rightarrow$}                                                                    & \multicolumn{1}{c|}{No}                          & \multicolumn{1}{c|}{1.073}                         & \multicolumn{1}{c|}{77.440}                                                               & \multicolumn{1}{c|}{\cellcolor[HTML]{FFFFFF}70.770 $\uparrow$}                                          & No          \\ \hline
10                              & 65275387                         & \multicolumn{1}{c|}{1.946}                         & 14.500                                                               & \multicolumn{1}{c|}{0.411}                         & \multicolumn{1}{c|}{85.710}                                                               & \multicolumn{1}{c|}{71.210 $\uparrow$}                                                                   & \multicolumn{1}{c|}{No}                          & \multicolumn{1}{c|}{0.076}                         & \multicolumn{1}{c|}{98.400}                                                               & \multicolumn{1}{c|}{\cellcolor[HTML]{FFFFFF}83.900 $\uparrow$}                                          & No          \\ \hline
\rowcolor[HTML]{D9D9D9} 
11                              & 54923573                         & \multicolumn{1}{c|}{\cellcolor[HTML]{D9D9D9}0.151} & 93.840                                                               & \multicolumn{1}{c|}{\cellcolor[HTML]{D9D9D9}0.177} & \multicolumn{1}{c|}{\cellcolor[HTML]{D9D9D9}92.770}                                       & \multicolumn{1}{c|}{\cellcolor[HTML]{D9D9D9}-1.070 $\downarrow$}                                           & \multicolumn{1}{c|}{\cellcolor[HTML]{D9D9D9}Yes} & \multicolumn{1}{c|}{\cellcolor[HTML]{D9D9D9}0.916} & \multicolumn{1}{c|}{\cellcolor[HTML]{D9D9D9}92.860}                                       & \multicolumn{1}{c|}{\cellcolor[HTML]{D9D9D9}-0.980 $\downarrow$}                                          & Yes         \\ \hline
12                              & 63027146                         & \multicolumn{1}{c|}{2.303}                         & 9.890                                                                & \multicolumn{1}{c|}{0.746}                         & \multicolumn{1}{c|}{74.240}                                                               & \multicolumn{1}{c|}{64.350 $\uparrow$}                                                                   & \multicolumn{1}{c|}{No}                          & \multicolumn{1}{c|}{0.721}                         & \multicolumn{1}{c|}{74.920}                                                               & \multicolumn{1}{c|}{\cellcolor[HTML]{FFFFFF}65.030 $\uparrow$}                                          & No          \\ \hline
13                              & 65659888                         & \multicolumn{1}{c|}{0.676}                         & 49.900                                                               & \multicolumn{1}{c|}{--}                            & \multicolumn{1}{c|}{--}                                                                   & \multicolumn{1}{c|}{--}                                                                       & \multicolumn{1}{c|}{--}                          & \multicolumn{1}{c|}{0.007}                         & \multicolumn{1}{c|}{93.010}                                                               & \multicolumn{1}{c|}{\cellcolor[HTML]{FFFFFF}43.110 $\uparrow$}                                          & No          \\ \hline
14                              & 58666904                         & \multicolumn{2}{c|}{Crash}                                                                                                & \multicolumn{1}{c|}{--}                            & \multicolumn{1}{c|}{--}                                                                   & \multicolumn{1}{c|}{--}                                                                       & \multicolumn{1}{c|}{--}                          & \multicolumn{1}{c|}{0.114}                         & \multicolumn{1}{c|}{72.500}                                                               & \multicolumn{1}{c|}{\cellcolor[HTML]{FFFFFF}72.500 $\uparrow$}                                          & No          \\ \hline
15                              & 55198221                         & \multicolumn{1}{c|}{0.998}                         & 60.320                                                               & \multicolumn{1}{c|}{0.960}                         & \multicolumn{1}{c|}{62.290}                                                               & \multicolumn{1}{c|}{1.970 $\uparrow$}                                                                    & \multicolumn{1}{c|}{No}                          & \multicolumn{1}{c|}{0.710}                         & \multicolumn{1}{c|}{72.320}                                                               & \multicolumn{1}{c|}{\cellcolor[HTML]{FFFFFF}12.000 $\uparrow$}                                          & No          \\ \hline
16                              & 55343875                         & \multicolumn{1}{c|}{0.064}                         & 97.240                                                               & \multicolumn{1}{c|}{0.039}                         & \multicolumn{1}{c|}{98.400}                                                               & \multicolumn{1}{c|}{1.160 $\uparrow$}                                                                    & \multicolumn{1}{c|}{No}                          & \multicolumn{1}{c|}{0.064}                         & \multicolumn{1}{c|}{97.800}                                                               & \multicolumn{1}{c|}{\cellcolor[HTML]{FFFFFF}0.560 $\uparrow$}                                           & No          \\ \hline
17                              & 38648195                         & \multicolumn{1}{c|}{0.201}                         & 47.810                                                               & \multicolumn{1}{c|}{0.839}                         & \multicolumn{1}{c|}{59.030}                                                               & \multicolumn{1}{c|}{11.220 $\uparrow$}                                                                   & \multicolumn{1}{c|}{No}                          & \multicolumn{1}{c|}{0.886}                         & \multicolumn{1}{c|}{58.030}                                                               & \multicolumn{1}{c|}{\cellcolor[HTML]{FFFFFF}10.220 $\uparrow$}                                          & No          \\ \hline
18                              & 48385830                         & \multicolumn{1}{c|}{nan}                           & 9.870                                                                & \multicolumn{1}{c|}{0.094}                         & \multicolumn{1}{c|}{10.110}                                                               & \multicolumn{1}{c|}{0.240 $\uparrow$}                                                                    & \multicolumn{1}{c|}{No}                          & \multicolumn{1}{c|}{1.086}                         & \multicolumn{1}{c|}{63.530}                                                               & \multicolumn{1}{c|}{\cellcolor[HTML]{FFFFFF}53.660 $\uparrow$}                                          & No          \\ \hline
19                              & 51930566                         & \multicolumn{1}{c|}{0.808}                         & 63.330                                                               & \multicolumn{1}{c|}{0.439}                         & \multicolumn{1}{c|}{76.890}                                                               & \multicolumn{1}{c|}{13.560 $\uparrow$}                                                                   & \multicolumn{1}{c|}{No}                          & \multicolumn{1}{c|}{0.811}                         & \multicolumn{1}{c|}{73.330}                                                               & \multicolumn{1}{c|}{\cellcolor[HTML]{FFFFFF}10.000 $\uparrow$}                                          & No          \\ \hline
20                              & 55328966                         & \multicolumn{1}{c|}{nan}                           & 9.860                                                                & \multicolumn{1}{c|}{0.136}                         & \multicolumn{1}{c|}{95.510}                                                               & \multicolumn{1}{c|}{85.650 $\uparrow$}                                                                   & \multicolumn{1}{c|}{No}                          & \multicolumn{1}{c|}{0.040}                         & \multicolumn{1}{c|}{98.710}                                                               & \multicolumn{1}{c|}{\cellcolor[HTML]{FFFFFF}88.850 $\uparrow$}                                          & No          \\ \hline
\rowcolor[HTML]{D9D9D9} 
21                              & 58609115                         & \multicolumn{1}{c|}{\cellcolor[HTML]{D9D9D9}0.016} & 99.780                                                               & \multicolumn{1}{c|}{\cellcolor[HTML]{D9D9D9}0.652} & \multicolumn{1}{c|}{\cellcolor[HTML]{D9D9D9}0.340}                                        & \multicolumn{1}{c|}{\cellcolor[HTML]{D9D9D9}-99.440 $\downarrow$}                                          & \multicolumn{1}{c|}{\cellcolor[HTML]{D9D9D9}Yes} & \multicolumn{1}{c|}{\cellcolor[HTML]{D9D9D9}0.623} & \multicolumn{1}{c|}{\cellcolor[HTML]{D9D9D9}0.400}                                        & \multicolumn{1}{c|}{\cellcolor[HTML]{D9D9D9}-99.380 $\downarrow$}                                         & Yes         \\ \hline
\rowcolor[HTML]{D9D9D9} 
22                              & 59278771                         & \multicolumn{1}{c|}{\cellcolor[HTML]{D9D9D9}0.107} & 97.330                                                               & \multicolumn{1}{c|}{\cellcolor[HTML]{D9D9D9}0.233} & \multicolumn{1}{c|}{\cellcolor[HTML]{D9D9D9}87.560}                                       & \multicolumn{1}{c|}{\cellcolor[HTML]{D9D9D9}-9.770 $\downarrow$}                                           & \multicolumn{1}{c|}{\cellcolor[HTML]{D9D9D9}Yes} & \multicolumn{1}{c|}{\cellcolor[HTML]{D9D9D9}0.477} & \multicolumn{1}{c|}{\cellcolor[HTML]{D9D9D9}88.000}                                       & \multicolumn{1}{c|}{\cellcolor[HTML]{D9D9D9}-9.330 $\downarrow$}                                          & Yes         \\ \hline
\end{tabular}

}
\begin{tablenotes}
 \centering
        \fontsize{7pt}{7pt}\selectfont
        \item The highlighted rows indicate programs where Theia's fix suggestions did not improve model performance.
        \item improvement \% = accuracy (in \%) after fix - accuracy (in \%) of buggy model.
        \item $\uparrow$ represents increase percentage, $\downarrow$ represents decrease percentage,
        $\rightarrow$ represents no change, and \textbf{--} indicates the model not supported yet.

	\end{tablenotes}
\label{tab:buggyMulticlass}

\end{table}
\begin{table}[h]
 \centering

    \captionof{table}{Comparison of Impact of Actionable Fixes by Theia and NeuraLint on Buggy Models Performance Designed for Binary Classification \& Regression Task.}
  \scalebox{0.685}{
\begin{tabular}{|c|c|cc|cccccccc|}
\hline
                                &                                  & \multicolumn{2}{c|}{}                                                                                                     & \multicolumn{8}{c|}{\textbf{\begin{tabular}[c]{@{}c@{}}Performance\\  After Fix\end{tabular}}}                                                                                                                                                                                                                                                                                                                                                                                                                                                                  \\ \cline{5-12} 
                                &                                  & \multicolumn{2}{c|}{\multirow{-2}{*}{\textbf{\begin{tabular}[c]{@{}c@{}}Performance\\ of Buggy Model\end{tabular}}}}      & \multicolumn{4}{c|}{\textbf{NeuraLint}}                                                                                                                                                                                                                                                          & \multicolumn{4}{c|}{\textbf{Theia}}                                                                                                                                                                                                                          \\ \cline{3-12} 
\multirow{-3}{*}{\textbf{SNo.}} & \multirow{-3}{*}{\textbf{SO \#}} & \multicolumn{1}{c|}{\textbf{Loss}}                 & \textbf{\begin{tabular}[c]{@{}c@{}}Accuracy\\  (in \%)\end{tabular}} & \multicolumn{1}{c|}{\textbf{Loss}}                 & \multicolumn{1}{c|}{\textbf{\begin{tabular}[c]{@{}c@{}}Accuracy\\  (in \%)\end{tabular}}} & \multicolumn{1}{c|}{\textbf{\begin{tabular}[c]{@{}c@{}}Improvement\\   (in \%)\end{tabular}}} & \multicolumn{1}{c|}{\textbf{FP}}                & \multicolumn{1}{c|}{\textbf{Loss}}                 & \multicolumn{1}{c|}{\textbf{\begin{tabular}[c]{@{}c@{}}Accuracy\\  (in \%)\end{tabular}}} & \multicolumn{1}{c|}{\textbf{\begin{tabular}[c]{@{}c@{}}Improvement\\   (in \%)\end{tabular}}} & \textbf{FP} \\ \hline
1                               & 58844149                         & \multicolumn{1}{c|}{7.645}                         & 49.860                                                               & \multicolumn{1}{c|}{0.359}                         & \multicolumn{1}{c|}{83.710}                                                               & \multicolumn{1}{c|}{33.850 $\uparrow$}                                                                   & \multicolumn{1}{c|}{No}                         & \multicolumn{1}{c|}{0.398}                         & \multicolumn{1}{c|}{81.520}                                                               & \multicolumn{1}{c|}{31.660 $\uparrow$}                                                                   & No          \\ \hline
2                               & 60261103                         & \multicolumn{1}{c|}{0.894}                         & 50.700                                                               & \multicolumn{1}{c|}{0.723}                         & \multicolumn{1}{c|}{49.600}                                                               & \multicolumn{1}{c|}{-1.100 $\downarrow$}                                                                   & \multicolumn{1}{c|}{Yes}                        & \multicolumn{1}{c|}{0.752}                         & \multicolumn{1}{c|}{50.900}                                                               & \multicolumn{1}{c|}{0.200 $\uparrow$}                                                                    & No          \\ \hline
3                               & 56914715                         & \multicolumn{1}{c|}{7.682}                         & 49.630                                                               & \multicolumn{1}{c|}{0.607}                         & \multicolumn{1}{c|}{74.250}                                                               & \multicolumn{1}{c|}{24.620 $\uparrow$}                                                                   & \multicolumn{1}{c|}{No}                         & \multicolumn{1}{c|}{0.136}                         & \multicolumn{1}{c|}{95.380}                                                               & \multicolumn{1}{c|}{45.750 $\uparrow$}                                                                   & No          \\ \hline
4                               & 60003876                         & \multicolumn{2}{c|}{Crash}                                                                                                & \multicolumn{1}{c|}{--}                            & \multicolumn{1}{c|}{--}                                                                   & \multicolumn{1}{c|}{--}                                                                       & \multicolumn{1}{c|}{--}                         & \multicolumn{1}{c|}{0.016}                         & \multicolumn{1}{c|}{91.700}                                                               & \multicolumn{1}{c|}{91.700 $\uparrow$}                                                                   & No          \\ \hline
5                               & 70428592                         & \multicolumn{2}{c|}{Crash}                                                                                                & \multicolumn{1}{c|}{--}                            & \multicolumn{1}{c|}{--}                                                                   & \multicolumn{1}{c|}{--}                                                                       & \multicolumn{1}{c|}{--}                         & \multicolumn{1}{c|}{0.003}                         & \multicolumn{1}{c|}{98.600}                                                               & \multicolumn{1}{c|}{98.600 $\uparrow$}                                                                   & No          \\ \hline
6                               & 40045159                         & \multicolumn{1}{c|}{0.490}                         & 72.880                                                               & \multicolumn{1}{c|}{0.485}                         & \multicolumn{1}{c|}{76.790}                                                               & \multicolumn{1}{c|}{3.910 $\uparrow$}                                                                    & \multicolumn{1}{c|}{No}                         & \multicolumn{1}{c|}{0.485}                         & \multicolumn{1}{c|}{76.750}                                                               & \multicolumn{1}{c|}{3.870 $\uparrow$}                                                                    & No          \\ \hline
7                               & 45378493                         & \multicolumn{1}{c|}{7.620}                         & 50.000                                                               & \multicolumn{1}{c|}{0.116}                         & \multicolumn{1}{c|}{97.000}                                                               & \multicolumn{1}{c|}{47.000 $\uparrow$}                                                                   & \multicolumn{1}{c|}{No}                         & \multicolumn{1}{c|}{0.073}                         & \multicolumn{1}{c|}{97.000}                                                               & \multicolumn{1}{c|}{47.000 $\uparrow$}                                                                   & No          \\ \hline
8                               & 51749207                         & \multicolumn{1}{c|}{7.655}                         & 49.800                                                               & \multicolumn{1}{c|}{0.008}                         & \multicolumn{1}{c|}{99.000}                                                               & \multicolumn{1}{c|}{49.200 $\uparrow$}                                                                   & \multicolumn{1}{c|}{No}                         & \multicolumn{1}{c|}{0.031}                         & \multicolumn{1}{c|}{99.000}                                                               & \multicolumn{1}{c|}{49.200 $\uparrow$}                                                                   & No          \\ \hline
9                               & 58844149                         & \multicolumn{1}{c|}{7.645}                         & 49.860                                                               & \multicolumn{1}{c|}{0.359}                         & \multicolumn{1}{c|}{83.710}                                                               & \multicolumn{1}{c|}{33.850 $\uparrow$}                                                                   & \multicolumn{1}{c|}{No}                         & \multicolumn{1}{c|}{0.398}                         & \multicolumn{1}{c|}{81.520}                                                               & \multicolumn{1}{c|}{31.660 $\uparrow$}                                                                   & No          \\ \hline
10                              & 31880720                         & \multicolumn{1}{c|}{7.660}                         & 50.000                                                               & \multicolumn{1}{c|}{0.003}                         & \multicolumn{1}{c|}{99.000}                                                               & \multicolumn{1}{c|}{49.000 $\uparrow$}                                                                   & \multicolumn{1}{c|}{No}                         & \multicolumn{1}{c|}{0.004}                         & \multicolumn{1}{c|}{99.810}                                                               & \multicolumn{1}{c|}{49.810 $\uparrow$}                                                                   & No          \\ \hline
11                              & 39525358                         & \multicolumn{1}{c|}{0.670}                         & 61.590                                                               & \multicolumn{1}{c|}{0.670}                         & \multicolumn{1}{c|}{61.590}                                                               & \multicolumn{1}{c|}{0.000 $\rightarrow$}                                                                    & \multicolumn{1}{c|}{No}                         & \multicolumn{1}{c|}{0.543}                         & \multicolumn{1}{c|}{92.310}                                                               & \multicolumn{1}{c|}{30.720 $\uparrow$}                                                                   & No          \\ \hline
12                              & 31627380                         & \multicolumn{1}{c|}{9.797}                         & 39.040                                                               & \multicolumn{1}{c|}{0.640}                         & \multicolumn{1}{c|}{68.260}                                                               & \multicolumn{1}{c|}{29.220 $\uparrow$}                                                                   & \multicolumn{1}{c|}{No}                         & \multicolumn{1}{c|}{0.428}                         & \multicolumn{1}{c|}{81.180}                                                               & \multicolumn{1}{c|}{42.140 $\uparrow$}                                                                   & No          \\ \hline
\rowcolor[HTML]{D9D9D9} 
13                              & 34673164                         & \multicolumn{1}{c|}{\cellcolor[HTML]{D9D9D9}0.128} & 77.780                                                               & \multicolumn{1}{c|}{\cellcolor[HTML]{D9D9D9}0.776} & \multicolumn{1}{c|}{\cellcolor[HTML]{D9D9D9}77.780}                                       & \multicolumn{1}{c|}{\cellcolor[HTML]{D9D9D9}0.000 $\rightarrow$}                                            & \multicolumn{1}{c|}{\cellcolor[HTML]{D9D9D9}No} & \multicolumn{1}{c|}{\cellcolor[HTML]{D9D9D9}0.450} & \multicolumn{1}{c|}{\cellcolor[HTML]{D9D9D9}77.780}                                       & \multicolumn{1}{c|}{\cellcolor[HTML]{D9D9D9}0.000 $\rightarrow$}                                            & No          \\ \hline
14                              & 34311586                         & \multicolumn{1}{c|}{0.667}                         & 33.300                                                               & \multicolumn{1}{c|}{0.689}                         & \multicolumn{1}{c|}{66.670}                                                               & \multicolumn{1}{c|}{33.370 $\uparrow$}                                                                   & \multicolumn{1}{c|}{No}                         & \multicolumn{1}{c|}{0.686}                         & \multicolumn{1}{c|}{66.670}                                                               & \multicolumn{1}{c|}{33.370 $\uparrow$}                                                                   & No          \\ \hline
15                              & 48221692                         & \multicolumn{1}{c|}{2288.030}                      & --                                                                   & \multicolumn{1}{c|}{-68722.021}                    & \multicolumn{1}{c|}{--}                                                                   & \multicolumn{1}{c|}{71010.051 $\uparrow$}                                                                & \multicolumn{1}{c|}{Yes}                        & \multicolumn{1}{c|}{916.983}                       & \multicolumn{1}{c|}{--}                                                                   & \multicolumn{1}{c|}{1371.047 $\downarrow$}                                                                 & No          \\ \hline
16                              & 48251943                         & \multicolumn{1}{c|}{736.928}                       & --                                                                   & \multicolumn{1}{c|}{736.928}                       & \multicolumn{1}{c|}{--}                                                                   & \multicolumn{1}{c|}{0.000 $\rightarrow$}                                                                    & \multicolumn{1}{c|}{No}                         & \multicolumn{1}{c|}{131.660}                       & \multicolumn{1}{c|}{--}                                                                   & \multicolumn{1}{c|}{605.268 $\downarrow$}                                                                  & No          \\ \hline
17                              & 48934338                         & \multicolumn{1}{c|}{1354.247}                      & --                                                                   & \multicolumn{1}{c|}{1354.247}                      & \multicolumn{1}{c|}{--}                                                                   & \multicolumn{1}{c|}{0.000 $\rightarrow$}                                                                    & \multicolumn{1}{c|}{No}                         & \multicolumn{1}{c|}{42.927}                        & \multicolumn{1}{c|}{--}                                                                   & \multicolumn{1}{c|}{1311.320 $\downarrow$}                                                                 & No          \\ \hline
\rowcolor[HTML]{D9D9D9} 
18                              & 44164749                         & \multicolumn{1}{c|}{\cellcolor[HTML]{D9D9D9}nan}   & 29.630                                                               & \multicolumn{1}{c|}{\cellcolor[HTML]{D9D9D9}nan}   & \multicolumn{1}{c|}{\cellcolor[HTML]{D9D9D9}29.630}                                       & \multicolumn{1}{c|}{\cellcolor[HTML]{D9D9D9}0.000 $\rightarrow$}                                            & \multicolumn{1}{c|}{\cellcolor[HTML]{D9D9D9}No} & \multicolumn{1}{c|}{\cellcolor[HTML]{D9D9D9}nan}   & \multicolumn{1}{c|}{\cellcolor[HTML]{D9D9D9}29.630}                                       & \multicolumn{1}{c|}{\cellcolor[HTML]{D9D9D9}0.000 $\rightarrow$}                                            & No          \\ \hline
\end{tabular}

}
\begin{tablenotes}
 \centering
        \fontsize{7pt}{7pt}\selectfont
        \item The highlighted rows indicate programs where Theia's fix suggestions did not improve model performance.
       \item improvement \% = accuracy (in \%) after fix - accuracy (in \%) of buggy model (for classification).
        \item improvement  = loss after fix - loss of buggy model (for regression).
        \item $\uparrow$ represents increase percentage, $\downarrow$ represents decrease percentage,
        $\rightarrow$ represents no change, and \textbf{--} indicates the model not supported yet.

	\end{tablenotes}
  \label{tab:buggyBinary}

\end{table}
\subsubsection{Evaluation on Binary Classification \& Regression Task}
To evaluate the effectiveness of \ourapproach on binary classification and regression tasks, we performed the evaluation on 18 buggy programs obtained from \sof in our benchmark. Table~\ref{tab:binary} reports the evaluation results of using \ourapproach and \nlint on these programs.
There are 13 programs for binary classification tasks in Table~\ref{tab:binary}. Most of the programs have \textit{LLM} bugs, \ourapproach successfully detected 12/12 bugs of this category by taking into account the dataset characteristics - number of classes.
Whereas, \nlint detected 2/12 bugs (SO\# 31627380, 34673164) in this category.
Both \ourapproach and \nlint support CNN program-specific bugs, therefore for SO\# 64188884, both the approaches detected bug in \textit{IDS} category.
For the regression task, there are 4 programs in Table~\ref{tab:binary},
\ourapproach successfully detected 4/4 bugs belonging to \textit{LLM, LOB} bug categories. On the other hand, \nlint was not able to find any of these bugs.
There is 1 program for the multi-label classification task, SO\# 44164749, both \ourapproach and \nlint failed to detect the bug in this program. 
We investigated the reason for it and found that for multi label classification, the mapping between the last layer activation function and loss function is different than the multiclass classification as discussed in~\ref{sec:Detection}. Therefore, in the verification rule, \textit{LLM}, there is a need to add a proper mapping.

\subsubsection{Evaluation of Actionable Fixes on Buggy DL Programs} 
\label{evaluationaf}
We investigated the impact of the fix suggestions provided by Theia on improving the performance of the buggy DL program after repair. 
To investigate this, the two authors manually addressed the bugs for each of the 40 programs in our benchmark,
following the line numbers and fix recommendations from \ourapproach and \nlint, and compared the results with the performance of the original buggy model.
If the fix does not improve the performance of the buggy model, we consider the fix suggestions as false alarms (FP). 
\ourapproach localizes the bug and provides 
developer hints at the potential solutions, whereas \nlint identifies the bug but does not provide guidance on potential solutions.
For instance, for inappropriate loss function, \ourapproach provides the message: ``Change loss function --> Use categorical\_crossentropy'', whereas, \nlint provides the message: ``Learner ==> The loss should be correctly defined and connected to the layer in accordance with its input conditions (\ie shape and type)-post\_activation''.
After fixing the bug, we rerun \ourapproach and \nlint on the modified DL program and repeat the process until no bugs are reported by both tools. 
The comparison of loss and accuracy of the original buggy program and the manually repaired program using actionable fixes from \ourapproach and \nlint are reported in Table~\ref{tab:buggyMulticlass} and Table~\ref{tab:buggyBinary}.
The results show that \ourapproach 
successfully provided actionable fixes, resulting in an average performance enhancement by 41\% in 34 out of 40 buggy DL programs. In contrast, the fix suggestions from \nlint led to an average performance improvement of 30\% in 19 out of 40 programs.
This highlights the effectiveness of our approach in detecting structural bugs that lead to suboptimal performance during training.

\begin{table}[h]
\centering
  \captionof{table}{Comparison of Impact of Actionable Fixes by Theia and NeuraLint on Normal Models Performance Designed for Multiclass Classification Task.}
  \scalebox{0.75}{
\begin{tabular}{|c|c|cc|cccccccc|}
\hline
                                &                                  & \multicolumn{2}{c|}{}                                                                                                     & \multicolumn{8}{c|}{\textbf{\begin{tabular}[c]{@{}c@{}}Performance\\  After Fix\end{tabular}}}                                                                                                                                                                                                                                                                                                                                                                                                                                                                   \\ \cline{5-12} 
                                &                                  & \multicolumn{2}{c|}{\multirow{-2}{*}{\textbf{\begin{tabular}[c]{@{}c@{}}Performance \\  of Normal Model\end{tabular}}}}  & \multicolumn{4}{c|}{\textbf{NeuraLint}}                                                                                                                                                                                                                                                           & \multicolumn{4}{c|}{\textbf{Theia}}                                                                                                                                                                                                                          \\ \cline{3-12} 
\multirow{-3}{*}{\textbf{SNo.}} & \multirow{-3}{*}{\textbf{SO \#}} & \multicolumn{1}{c|}{\textbf{Loss}}                 & \textbf{\begin{tabular}[c]{@{}c@{}}Accuracy\\  (in \%)\end{tabular}} & \multicolumn{1}{c|}{\textbf{Loss}}                 & \multicolumn{1}{c|}{\textbf{\begin{tabular}[c]{@{}c@{}}Accuracy\\  (in \%)\end{tabular}}} & \multicolumn{1}{c|}{\textbf{\begin{tabular}[c]{@{}c@{}}Improvement\\   (in \%)\end{tabular}}} & \multicolumn{1}{c|}{\textbf{FP}}                 & \multicolumn{1}{c|}{\textbf{Loss}}                 & \multicolumn{1}{c|}{\textbf{\begin{tabular}[c]{@{}c@{}}Accuracy\\  (in \%)\end{tabular}}} & \multicolumn{1}{c|}{\textbf{\begin{tabular}[c]{@{}c@{}}Improvement\\   (in \%)\end{tabular}}} & \textbf{FP} \\ \hline
\rowcolor[HTML]{D9D9D9} 
1                               & 64522751                         & \multicolumn{1}{c|}{\cellcolor[HTML]{D9D9D9}0.298} & 89.250                                                               & \multicolumn{1}{c|}{\cellcolor[HTML]{D9D9D9}0.282} & \multicolumn{1}{c|}{\cellcolor[HTML]{D9D9D9}89.740}                                       & \multicolumn{1}{c|}{\cellcolor[HTML]{D9D9D9}0.490 $\uparrow$}                                            & \multicolumn{1}{c|}{\cellcolor[HTML]{D9D9D9}No}  & \multicolumn{1}{c|}{\cellcolor[HTML]{D9D9D9}0.492} & \multicolumn{1}{c|}{\cellcolor[HTML]{D9D9D9}82.920}                                       & \multicolumn{1}{c|}{\cellcolor[HTML]{D9D9D9}-6.330 $\downarrow$}                                           & Yes         \\ \hline
2                               & 50079585                         & \multicolumn{1}{c|}{0.511}                         & 79.100                                                               & \multicolumn{1}{c|}{0.816}                         & \multicolumn{1}{c|}{63.240}                                                               & \multicolumn{1}{c|}{-15.860 $\downarrow$}                                                                  & \multicolumn{1}{c|}{Yes}                         & \multicolumn{1}{c|}{0.526}                         & \multicolumn{1}{c|}{79.830}                                                               & \multicolumn{1}{c|}{0.730 $\uparrow$}                                                                    & No          \\ \hline
3                               & 47272383                         & \multicolumn{1}{c|}{0.480}                         & 79.600                                                               & \multicolumn{1}{c|}{1.763}                         & \multicolumn{1}{c|}{20.450}                                                               & \multicolumn{1}{c|}{-59.150 $\downarrow$}                                                                  & \multicolumn{1}{c|}{Yes}                         & \multicolumn{1}{c|}{0.392}                         & \multicolumn{1}{c|}{87.340}                                                               & \multicolumn{1}{c|}{7.740 $\uparrow$}                                                                    & No          \\ \hline
4                               & 51118032                         & \multicolumn{1}{c|}{0.599}                         & 79.580                                                               & \multicolumn{1}{c|}{0.599}                         & \multicolumn{1}{c|}{79.580}                                                               & \multicolumn{1}{c|}{0.000 $\rightarrow$}                                                                    & \multicolumn{1}{c|}{No}                          & \multicolumn{1}{c|}{0.315}                         & \multicolumn{1}{c|}{88.690}                                                               & \multicolumn{1}{c|}{9.110 $\uparrow$}                                                                    & No          \\ \hline
\rowcolor[HTML]{D9D9D9} 
5                               & 37229086                         & \multicolumn{1}{c|}{\cellcolor[HTML]{D9D9D9}0.314} & 88.530                                                               & \multicolumn{1}{c|}{\cellcolor[HTML]{D9D9D9}0.237} & \multicolumn{1}{c|}{\cellcolor[HTML]{D9D9D9}91.550}                                       & \multicolumn{1}{c|}{\cellcolor[HTML]{D9D9D9}3.020 $\uparrow$}                                            & \multicolumn{1}{c|}{\cellcolor[HTML]{D9D9D9}No}  & \multicolumn{1}{c|}{\cellcolor[HTML]{D9D9D9}0.429} & \multicolumn{1}{c|}{\cellcolor[HTML]{D9D9D9}84.790}                                       & \multicolumn{1}{c|}{\cellcolor[HTML]{D9D9D9}-3.740 $\downarrow$}                                           & Yes         \\ \hline
6                               & 48594888                         & \multicolumn{1}{c|}{0.424}                         & 85.810                                                               & \multicolumn{1}{c|}{0.424}                         & \multicolumn{1}{c|}{85.810}                                                               & \multicolumn{1}{c|}{0.000 $\rightarrow$}                                                                    & \multicolumn{1}{c|}{No}                          & \multicolumn{1}{c|}{0.333}                         & \multicolumn{1}{c|}{88.270}                                                               & \multicolumn{1}{c|}{2.460}                                                                    & No          \\ \hline
7                               & 59325381                         & \multicolumn{1}{c|}{0.050}                         & 98.830                                                               & \multicolumn{1}{c|}{0.040}                         & \multicolumn{1}{c|}{98.950}                                                               & \multicolumn{1}{c|}{0.120 $\uparrow$}                                                                    & \multicolumn{1}{c|}{No}                          & \multicolumn{1}{c|}{0.024}                         & \multicolumn{1}{c|}{99.270}                                                               & \multicolumn{1}{c|}{0.440 $\uparrow$}                                                                    & No          \\ \hline
8                               & 64188884                         & \multicolumn{1}{c|}{0.899}                         & 65.760                                                               & \multicolumn{1}{c|}{0.899}                         & \multicolumn{1}{c|}{65.760}                                                               & \multicolumn{1}{c|}{0.000 $\rightarrow$}                                                                    & \multicolumn{1}{c|}{No}                          & \multicolumn{1}{c|}{0.760}                         & \multicolumn{1}{c|}{71.440}                                                               & \multicolumn{1}{c|}{5.680 $\uparrow$}                                                                    & No          \\ \hline
9                               & 70554413                         & \multicolumn{1}{c|}{1.072}                         & 76.500                                                               & \multicolumn{1}{c|}{2.770}                         & \multicolumn{1}{c|}{7.440}                                                                & \multicolumn{1}{c|}{-69.060 $\downarrow$}                                                                  & \multicolumn{1}{c|}{Yes}                         & \multicolumn{1}{c|}{1.098}                         & \multicolumn{1}{c|}{78.330}                                                               & \multicolumn{1}{c|}{1.830 $\uparrow$}                                                                    & No          \\ \hline
\rowcolor[HTML]{D9D9D9} 
10                              & 65275387                         & \multicolumn{1}{c|}{\cellcolor[HTML]{D9D9D9}1.946} & 84.500                                                               & \multicolumn{1}{c|}{\cellcolor[HTML]{D9D9D9}1.854} & \multicolumn{1}{c|}{\cellcolor[HTML]{D9D9D9}16.700}                                       & \multicolumn{1}{c|}{\cellcolor[HTML]{D9D9D9}-67.800 $\downarrow$}                                          & \multicolumn{1}{c|}{\cellcolor[HTML]{D9D9D9}Yes} & \multicolumn{1}{c|}{\cellcolor[HTML]{D9D9D9}1.854} & \multicolumn{1}{c|}{\cellcolor[HTML]{D9D9D9}20.000}                                       & \multicolumn{1}{c|}{\cellcolor[HTML]{D9D9D9}-64.500 $\downarrow$}                                          & Yes         \\ \hline
11                              & 54923573                         & \multicolumn{1}{c|}{0.160}                         & 94.000                                                               & \multicolumn{1}{c|}{0.233}                         & \multicolumn{1}{c|}{90.150}                                                               & \multicolumn{1}{c|}{-3.850 $\downarrow$}                                                                   & \multicolumn{1}{c|}{Yes}                         & \multicolumn{1}{c|}{0.194}                         & \multicolumn{1}{c|}{94.100}                                                               & \multicolumn{1}{c|}{0.100 $\uparrow$}                                                                    & No          \\ \hline
\rowcolor[HTML]{D9D9D9} 
12                              & 63027146                         & \multicolumn{1}{c|}{\cellcolor[HTML]{D9D9D9}0.253} & 91.000                                                               & \multicolumn{1}{c|}{\cellcolor[HTML]{D9D9D9}0.414} & \multicolumn{1}{c|}{\cellcolor[HTML]{D9D9D9}85.390}                                       & \multicolumn{1}{c|}{\cellcolor[HTML]{D9D9D9}-5.610 $\downarrow$}                                           & \multicolumn{1}{c|}{\cellcolor[HTML]{D9D9D9}Yes} & \multicolumn{1}{c|}{\cellcolor[HTML]{D9D9D9}0.495} & \multicolumn{1}{c|}{\cellcolor[HTML]{D9D9D9}82.840}                                       & \multicolumn{1}{c|}{\cellcolor[HTML]{D9D9D9}-8.160 $\downarrow$}                                           & Yes         \\ \hline
13                              & 65659888                         & \multicolumn{1}{c|}{0.076}                         & 90.100                                                               & \multicolumn{1}{c|}{--}                            & \multicolumn{1}{c|}{--}                                                                   & \multicolumn{1}{c|}{--}                                                                       & \multicolumn{1}{c|}{--}                          & \multicolumn{1}{c|}{0.004}                         & \multicolumn{1}{c|}{94.410}                                                               & \multicolumn{1}{c|}{4.310 $\uparrow$}                                                                    & No          \\ \hline
14                              & 58666904                         & \multicolumn{1}{c|}{0.112}                         & 70.100                                                               & \multicolumn{1}{c|}{--}                            & \multicolumn{1}{c|}{--}                                                                   & \multicolumn{1}{c|}{--}                                                                       & \multicolumn{1}{c|}{--}                          & \multicolumn{1}{c|}{0.114}                         & \multicolumn{1}{c|}{72.500}                                                               & \multicolumn{1}{c|}{2.400 $\uparrow$}                                                                    & No          \\ \hline
15                              & 55198221                         & \multicolumn{1}{c|}{0.935}                         & 63.840                                                               & \multicolumn{1}{c|}{0.935}                         & \multicolumn{1}{c|}{63.840}                                                               & \multicolumn{1}{c|}{0.000 $\rightarrow$}                                                                    & \multicolumn{1}{c|}{No}                          & \multicolumn{1}{c|}{0.807}                         & \multicolumn{1}{c|}{68.570}                                                               & \multicolumn{1}{c|}{4.730 $\uparrow$}                                                                    & No          \\ \hline
16                              & 55343875                         & \multicolumn{1}{c|}{0.051}                         & 98.290                                                               & \multicolumn{1}{c|}{0.051}                         & \multicolumn{1}{c|}{98.290}                                                               & \multicolumn{1}{c|}{0.000 $\rightarrow$}                                                                    & \multicolumn{1}{c|}{No}                          & \multicolumn{1}{c|}{0.022}                         & \multicolumn{1}{c|}{99.760}                                                               & \multicolumn{1}{c|}{1.470 $\uparrow$}                                                                    & No          \\ \hline
17                              & 38648195                         & \multicolumn{1}{c|}{0.727}                         & 68.190                                                               & \multicolumn{1}{c|}{0.727}                         & \multicolumn{1}{c|}{68.190}                                                               & \multicolumn{1}{c|}{0.000 $\rightarrow$}                                                                    & \multicolumn{1}{c|}{No}                          & \multicolumn{1}{c|}{0.728}                         & \multicolumn{1}{c|}{68.400}                                                               & \multicolumn{1}{c|}{0.210 $\uparrow$}                                                                    & No          \\ \hline
\rowcolor[HTML]{D9D9D9} 
18                              & 48385830                         & \multicolumn{1}{c|}{\cellcolor[HTML]{D9D9D9}0.290} & 91.200                                                               & \multicolumn{1}{c|}{\cellcolor[HTML]{D9D9D9}0.290} & \multicolumn{1}{c|}{\cellcolor[HTML]{D9D9D9}91.200}                                       & \multicolumn{1}{c|}{\cellcolor[HTML]{D9D9D9}0.000 $\rightarrow$}                                            & \multicolumn{1}{c|}{\cellcolor[HTML]{D9D9D9}No}  & \multicolumn{1}{c|}{\cellcolor[HTML]{D9D9D9}0.723} & \multicolumn{1}{c|}{\cellcolor[HTML]{D9D9D9}76.760}                                       & \multicolumn{1}{c|}{\cellcolor[HTML]{D9D9D9}-14.440 $\downarrow$}                                          & Yes         \\ \hline
\rowcolor[HTML]{D9D9D9} 
19                              & 51930566                         & \multicolumn{1}{c|}{\cellcolor[HTML]{D9D9D9}0.454} & 91.670                                                               & \multicolumn{1}{c|}{\cellcolor[HTML]{D9D9D9}0.454} & \multicolumn{1}{c|}{\cellcolor[HTML]{D9D9D9}91.670}                                       & \multicolumn{1}{c|}{\cellcolor[HTML]{D9D9D9}0.000 $\rightarrow$}                                            & \multicolumn{1}{c|}{\cellcolor[HTML]{D9D9D9}No}  & \multicolumn{1}{c|}{\cellcolor[HTML]{D9D9D9}0.698} & \multicolumn{1}{c|}{\cellcolor[HTML]{D9D9D9}81.670}                                       & \multicolumn{1}{c|}{\cellcolor[HTML]{D9D9D9}-10.000 $\downarrow$}                                          & Yes         \\ \hline
20                              & 55328966                         & \multicolumn{1}{c|}{0.008}                         & 99.770                                                               & \multicolumn{1}{c|}{0.008}                         & \multicolumn{1}{c|}{99.770}                                                               & \multicolumn{1}{c|}{0.000 $\rightarrow$}                                                                    & \multicolumn{1}{c|}{No}                          & \multicolumn{1}{c|}{0.030}                         & \multicolumn{1}{c|}{99.770}                                                               & \multicolumn{1}{c|}{0.000 $\rightarrow$}                                                                    & No          \\ \hline
\rowcolor[HTML]{D9D9D9} 
21                              & 58609115                         & \multicolumn{1}{c|}{\cellcolor[HTML]{D9D9D9}0.070} & 99.800                                                               & \multicolumn{1}{c|}{\cellcolor[HTML]{D9D9D9}0.776} & \multicolumn{1}{c|}{\cellcolor[HTML]{D9D9D9}0.170}                                        & \multicolumn{1}{c|}{\cellcolor[HTML]{D9D9D9}-99.630 $\downarrow$}                                          & \multicolumn{1}{c|}{\cellcolor[HTML]{D9D9D9}Yes} & \multicolumn{1}{c|}{\cellcolor[HTML]{D9D9D9}0.633} & \multicolumn{1}{c|}{\cellcolor[HTML]{D9D9D9}0.500}                                        & \multicolumn{1}{c|}{\cellcolor[HTML]{D9D9D9}-99.300 $\downarrow$}                                          & Yes         \\ \hline
22                              & 59278771                         & \multicolumn{1}{c|}{0.059}                         & 96.670                                                               & \multicolumn{1}{c|}{0.059}                         & \multicolumn{1}{c|}{96.670}                                                               & \multicolumn{1}{c|}{0.000 $\rightarrow$}                                                                    & \multicolumn{1}{c|}{No}                          & \multicolumn{1}{c|}{0.106}                         & \multicolumn{1}{c|}{96.670}                                                               & \multicolumn{1}{c|}{0.000}                                                                    & No          \\ \hline

\end{tabular}

}
\begin{tablenotes}
 \centering
         \fontsize{7pt}{7pt}\selectfont
         \item The highlighted rows indicate the false positives reported by \ourapproach.
         \item improvement \% = accuracy (in \%) after fix - accuracy (in \%) of buggy model (for classification).
        \item improvement  = loss after fix - loss of buggy model (for regression).
        \item $\uparrow$ represents increase percentage, $\downarrow$ represents decrease percentage,
        $\rightarrow$ represents no change, and \textbf{--} indicates the model not supported yet.

	\end{tablenotes}
\label{tab:correctMulticlass}

\end{table}
\subsubsection{Evaluation on Normal Programs}
We conducted a more thorough investigation into the effects of applying \ourapproach on normal programs, aiming to investigate any instances of false alarms in these programs.
As shown in Table~\ref{tab:patchMulticlass} and Table~\ref{tab:patchBinary}, the patches/fixes suggested by \sof users successfully resolved the bugs present in all 40 programs in our benchmark, resulting in improved performance.
Therefore, we utilized these patches to create a benchmark of 40 normal programs. These programs are available in our repository~\cite{myRepo}. 
We evaluated the impact of fix suggestions provided by Theia on these 40 normal programs.
We followed the same procedure as described in Section~\ref{evaluationaf}.
Two authors manually addressed the bugs for each of the 40 normal programs,
following the line numbers and fix recommendations from \ourapproach and \nlint and compared it with the performance of the normal model.
If the fix does not improve the performance of the model, we consider the fix suggestions as false alarms (FP). 
The impact on the performance of the normal programs after applying patches is shown in Table~\ref{tab:correctMulticlass} and Table~\ref{tab:correctBinary}.
On normal programs both \ourapproach (average performance improvement of 6\%) and \nlint (average performance improvement of 4\%) demonstrated a comparable performance on 40 programs, leading to performance improvements in 28 out of the 40 normal programs. 
Both tools negatively impacted the performance of 12 programs, resulting in 12 false alarms (FP). 
We investigated the reason for false alarms produced by \ourapproach in 12 out of 40 programs.
\ourapproach suggests to add Batch Normalization and Dropout layers after convolution and dense layers (Rules - \textit{MNL} and \textit{MRD}).
We observed that for less complex models, the addition of these layers after each convolution and dense layer leads to excessive regularization, thereby compromising the performance of these models.

\begin{table}[h]
\centering
    \captionof{table}{Comparison of Impact of Actionable Fixes by Theia and NeuraLint on Normal Models Performance Designed for Binary Classification \& Regression Task.}
  \scalebox{0.68}{
\begin{tabular}{|c|c|cc|cccccccc|}
\hline
                                &                                  & \multicolumn{2}{c|}{}                                                                                                           & \multicolumn{8}{c|}{\textbf{\begin{tabular}[c]{@{}c@{}}Performance\\  After Fix\end{tabular}}}                                                                                                                                                                                                                                                                                                                                                                                                                                                                           \\ \cline{5-12} 
                                &                                  & \multicolumn{2}{c|}{\multirow{-2}{*}{\textbf{\begin{tabular}[c]{@{}c@{}}Performance \\  of Normal Model\end{tabular}}}}        & \multicolumn{4}{c|}{\textbf{NeuraLint}}                                                                                                                                                                                                                                                                 & \multicolumn{4}{c|}{\textbf{Theia}}                                                                                                                                                                                                                            \\ \cline{3-12} 
\multirow{-3}{*}{\textbf{SNo.}} & \multirow{-3}{*}{\textbf{SO \#}} & \multicolumn{1}{c|}{\textbf{Loss}}                       & \textbf{\begin{tabular}[c]{@{}c@{}}Accuracy\\  (in \%)\end{tabular}} & \multicolumn{1}{c|}{\textbf{Loss}}                       & \multicolumn{1}{c|}{\textbf{\begin{tabular}[c]{@{}c@{}}Accuracy\\  (in \%)\end{tabular}}} & \multicolumn{1}{c|}{\textbf{\begin{tabular}[c]{@{}c@{}}Improvement\\   (in \%)\end{tabular}}} & \multicolumn{1}{c|}{\textbf{FP}}                 & \multicolumn{1}{c|}{\textbf{Loss}}                   & \multicolumn{1}{c|}{\textbf{\begin{tabular}[c]{@{}c@{}}Accuracy\\  (in \%)\end{tabular}}} & \multicolumn{1}{c|}{\textbf{\begin{tabular}[c]{@{}c@{}}Improvement\\   (in \%)\end{tabular}}} & \textbf{FP} \\ \hline
1                               & 58844149                         & \multicolumn{1}{c|}{0.485}                               & 76.030                                                               & \multicolumn{1}{c|}{0.350}                               & \multicolumn{1}{c|}{84.210}                                                               & \multicolumn{1}{c|}{8.180 $\uparrow$}                                                                    & \multicolumn{1}{c|}{No}                          & \multicolumn{1}{c|}{0.392}                           & \multicolumn{1}{c|}{81.910}                                                               & \multicolumn{1}{c|}{5.880 $\uparrow$}                                                                    & No          \\ \hline
\rowcolor[HTML]{D9D9D9} 
2                               & 60261103                         & \multicolumn{1}{c|}{\cellcolor[HTML]{D9D9D9}0.100}       & 96.400                                                               & \multicolumn{1}{c|}{\cellcolor[HTML]{D9D9D9}0.732}       & \multicolumn{1}{c|}{\cellcolor[HTML]{D9D9D9}51.100}                                       & \multicolumn{1}{c|}{\cellcolor[HTML]{D9D9D9}-45.300 $\downarrow$}                                          & \multicolumn{1}{c|}{\cellcolor[HTML]{D9D9D9}Yes} & \multicolumn{1}{c|}{\cellcolor[HTML]{D9D9D9}0.755}   & \multicolumn{1}{c|}{\cellcolor[HTML]{D9D9D9}52.100}                                       & \multicolumn{1}{c|}{\cellcolor[HTML]{D9D9D9}-44.300 $\downarrow$}                                          & Yes         \\ \hline
3                               & 56914715                         & \multicolumn{1}{c|}{0.490}                               & 84.750                                                               & \multicolumn{1}{c|}{0.490}                               & \multicolumn{1}{c|}{84.750}                                                               & \multicolumn{1}{c|}{0.000 $\rightarrow$}                                                                    & \multicolumn{1}{c|}{No}                          & \multicolumn{1}{c|}{0.154}                           & \multicolumn{1}{c|}{95.880}                                                               & \multicolumn{1}{c|}{11.130 $\uparrow$}                                                                   & No          \\ \hline
4                               & 60003876                         & \multicolumn{1}{c|}{0.938}                               & 50.000                                                               & \multicolumn{1}{c|}{--}                                  & \multicolumn{1}{c|}{--}                                                                   & \multicolumn{1}{c|}{--}                                                                       & \multicolumn{1}{c|}{--}                          & \multicolumn{1}{c|}{0.014}                           & \multicolumn{1}{c|}{92.100}                                                               & \multicolumn{1}{c|}{42.100 $\uparrow$}                                                                   & No          \\ \hline
\rowcolor[HTML]{D9D9D9} 
5                               & 70428592                         & \multicolumn{1}{c|}{\cellcolor[HTML]{D9D9D9}0.000}       & 98.100                                                               & \multicolumn{1}{c|}{\cellcolor[HTML]{D9D9D9}--}          & \multicolumn{1}{c|}{\cellcolor[HTML]{D9D9D9}--}                                           & \multicolumn{1}{c|}{\cellcolor[HTML]{D9D9D9}--}                                               & \multicolumn{1}{c|}{\cellcolor[HTML]{D9D9D9}--}  & \multicolumn{1}{c|}{\cellcolor[HTML]{D9D9D9}0.494}   & \multicolumn{1}{c|}{\cellcolor[HTML]{D9D9D9}81.200}                                       & \multicolumn{1}{c|}{\cellcolor[HTML]{D9D9D9}-16.900 $\downarrow$}                                          & Yes         \\ \hline
6                               & 40045159                         & \multicolumn{1}{c|}{0.441}                               & 80.100                                                               & \multicolumn{1}{c|}{0.441}                               & \multicolumn{1}{c|}{80.100}                                                               & \multicolumn{1}{c|}{0.000 $\rightarrow$}                                                                    & \multicolumn{1}{c|}{No}                          & \multicolumn{1}{c|}{0.467}                           & \multicolumn{1}{c|}{80.400}                                                               & \multicolumn{1}{c|}{0.300 $\uparrow$}                                                                    & No          \\ \hline
7                               & 45378493                         & \multicolumn{1}{c|}{0.076}                               & 99.000                                                               & \multicolumn{1}{c|}{0.113}                               & \multicolumn{1}{c|}{98.000}                                                               & \multicolumn{1}{c|}{-1.000 $\downarrow$}                                                                   & \multicolumn{1}{c|}{Yes}                         & \multicolumn{1}{c|}{0.069}                           & \multicolumn{1}{c|}{99.100}                                                               & \multicolumn{1}{c|}{0.100 $\uparrow$}                                                                    & No          \\ \hline
8                               & 51749207                         & \multicolumn{1}{c|}{0.011}                               & 99.600                                                               & \multicolumn{1}{c|}{0.011}                               & \multicolumn{1}{c|}{99.600}                                                               & \multicolumn{1}{c|}{0.000 $\rightarrow$}                                                                    & \multicolumn{1}{c|}{No}                          & \multicolumn{1}{c|}{0.013}                           & \multicolumn{1}{c|}{99.600}                                                               & \multicolumn{1}{c|}{0.000 $\rightarrow$}                                                                    & No          \\ \hline
9                               & 58844149                         & \multicolumn{1}{c|}{0.485}                               & 76.030                                                               & \multicolumn{1}{c|}{0.350}                               & \multicolumn{1}{c|}{84.210}                                                               & \multicolumn{1}{c|}{8.180 $\uparrow$}                                                                    & \multicolumn{1}{c|}{No}                          & \multicolumn{1}{c|}{0.392}                           & \multicolumn{1}{c|}{81.910}                                                               & \multicolumn{1}{c|}{5.880 $\uparrow$}                                                                    & No          \\ \hline
10                              & 31880720                         & \multicolumn{1}{c|}{0.005}                               & 99.900                                                               & \multicolumn{1}{c|}{0.003}                               & \multicolumn{1}{c|}{99.000}                                                               & \multicolumn{1}{c|}{-0.900 $\downarrow$}                                                                   & \multicolumn{1}{c|}{Yes}                         & \multicolumn{1}{c|}{0.027}                           & \multicolumn{1}{c|}{99.900}                                                               & \multicolumn{1}{c|}{0.000 $\rightarrow$}                                                                    & No          \\ \hline
11                              & 39525358                         & \multicolumn{1}{c|}{0.575}                               & 92.310                                                                & \multicolumn{1}{c|}{0.575}                               & \multicolumn{1}{c|}{92.310}                                                               & \multicolumn{1}{c|}{0.000 $\rightarrow$}                                                                   & \multicolumn{1}{c|}{No}                          & \multicolumn{1}{c|}{0.527}                           & \multicolumn{1}{c|}{99.000}                                                               & \multicolumn{1}{c|}{6.690 $\uparrow$}                                                                   & No          \\ \hline
12                              & 31627380                         & \multicolumn{1}{c|}{0.643}                               & 68.120                                                               & \multicolumn{1}{c|}{0.643}                               & \multicolumn{1}{c|}{68.120}                                                               & \multicolumn{1}{c|}{0.000 $\rightarrow$}                                                                    & \multicolumn{1}{c|}{No}                          & \multicolumn{1}{c|}{0.423}                           & \multicolumn{1}{c|}{82.580}                                                               & \multicolumn{1}{c|}{14.460 $\uparrow$}                                                                   & No          \\ \hline
13                              & 34673164                         & \multicolumn{1}{c|}{0.422}                               & 88.890                                                               & \multicolumn{1}{c|}{3.931}                               & \multicolumn{1}{c|}{77.780}                                                               & \multicolumn{1}{c|}{-11.110 $\downarrow$}                                                                  & \multicolumn{1}{c|}{Yes}                         & \multicolumn{1}{c|}{0.479}                           & \multicolumn{1}{c|}{88.900}                                                               & \multicolumn{1}{c|}{0.010 $\uparrow$}                                                                    & No          \\ \hline
14                              & 34311586                         & \multicolumn{1}{c|}{0.684}                               & 66.670                                                               & \multicolumn{1}{c|}{0.679}                               & \multicolumn{1}{c|}{66.670}                                                               & \multicolumn{1}{c|}{0.000 $\rightarrow$}                                                                    & \multicolumn{1}{c|}{No}                          & \multicolumn{1}{c|}{0.680}                           & \multicolumn{1}{c|}{66.670}                                                               & \multicolumn{1}{c|}{0.000 $\rightarrow$}                                                                    & No          \\ \hline
\rowcolor[HTML]{D9D9D9} 
15                              & 48221692                         & \multicolumn{1}{c|}{\cellcolor[HTML]{D9D9D9}95.283}      & --                                                                   & \multicolumn{1}{c|}{\cellcolor[HTML]{D9D9D9}95.283}      & \multicolumn{1}{c|}{\cellcolor[HTML]{D9D9D9}--}                                           & \multicolumn{1}{c|}{\cellcolor[HTML]{D9D9D9}0.000 $\rightarrow$}                                            & \multicolumn{1}{c|}{\cellcolor[HTML]{D9D9D9}No}  & \multicolumn{1}{c|}{\cellcolor[HTML]{D9D9D9}917.720} & \multicolumn{1}{c|}{\cellcolor[HTML]{D9D9D9}--}                                           & \multicolumn{1}{c|}{\cellcolor[HTML]{D9D9D9}-822.437 $\uparrow$}                                         & Yes         \\ \hline
\rowcolor[HTML]{D9D9D9} 
16                              & 48251943                         & \multicolumn{1}{c|}{\cellcolor[HTML]{D9D9D9}0.000018446} & --                                                                   & \multicolumn{1}{c|}{\cellcolor[HTML]{D9D9D9}0.000018446} & \multicolumn{1}{c|}{\cellcolor[HTML]{D9D9D9}--}                                           & \multicolumn{1}{c|}{\cellcolor[HTML]{D9D9D9}0.000 $\rightarrow$}                                            & \multicolumn{1}{c|}{\cellcolor[HTML]{D9D9D9}No}  & \multicolumn{1}{c|}{\cellcolor[HTML]{D9D9D9}131.660} & \multicolumn{1}{c|}{\cellcolor[HTML]{D9D9D9}--}                                           & \multicolumn{1}{c|}{\cellcolor[HTML]{D9D9D9}-131.660 $\uparrow$}                                         & Yes         \\ \hline
17                              & 48934338                         & \multicolumn{1}{c|}{248.703}                             & --                                                                   & \multicolumn{1}{c|}{248.703}                             & \multicolumn{1}{c|}{--}                                                                   & \multicolumn{1}{c|}{0.000 $\rightarrow$}                                                                    & \multicolumn{1}{c|}{No}                          & \multicolumn{1}{c|}{76.370}                          & \multicolumn{1}{c|}{--}                                                                   & \multicolumn{1}{c|}{172.333 $\downarrow$}                                                                  & No          \\ \hline
\rowcolor[HTML]{D9D9D9} 
18                              & 44164749                         & \multicolumn{1}{c|}{\cellcolor[HTML]{D9D9D9}0.449}       & 79.210                                                               & \multicolumn{1}{c|}{\cellcolor[HTML]{D9D9D9}nan}         & \multicolumn{1}{c|}{\cellcolor[HTML]{D9D9D9}29.630}                                       & \multicolumn{1}{c|}{\cellcolor[HTML]{D9D9D9}-49.580 $\downarrow$}                                          & \multicolumn{1}{c|}{\cellcolor[HTML]{D9D9D9}Yes} & \multicolumn{1}{c|}{\cellcolor[HTML]{D9D9D9}nan}     & \multicolumn{1}{c|}{\cellcolor[HTML]{D9D9D9}29.630}                                       & \multicolumn{1}{c|}{\cellcolor[HTML]{D9D9D9}-49.580 $\downarrow$}                                          & Yes         \\ \hline

\end{tabular}

}
\begin{tablenotes}
 \centering
        \fontsize{7pt}{7pt}\selectfont
        \item The highlighted rows indicate the false positives reported by \ourapproach.
         \item improvement \% = accuracy (in \%) after fix - accuracy (in \%) of buggy model (for classification).
        \item improvement  = loss after fix - loss of buggy model (for regression).
        \item $\uparrow$ represents increase percentage, $\downarrow$ represents decrease percentage,
        $\rightarrow$ represents no change, and \textbf{--} indicates the model not supported yet.
	\end{tablenotes}
 \label{tab:correctBinary}
\end{table}

\begin{table}[h]
\small
\caption{Comparison of Bugs Localized by Theia and NeuraLint Across Different Bug Categories.}
\begin{tabular}{|c|c|c|c|}
\hline
\textbf{Bug Categories} & \multicolumn{1}{c|}{\textbf{Total Bugs}} & \multicolumn{1}{c|}{\textbf{Theia}} & \multicolumn{1}{c|}{\textbf{NeuraLint}} \\ \hline
\rowcolor[HTML]{D0CECE}    \textbf{LLM}   & 35                                       & 33                                  & 11                                      \\ \hline
\textbf{CNL}   & 5                                        & 4                                   & 1                                       \\ \hline
\rowcolor[HTML]{D0CECE}   \textbf{ICL}   & 5                                        & 5                                   & --                                      \\ \hline
\textbf{IDN}   & 5                                        & 5                                   & --                                      \\ \hline
\rowcolor[HTML]{D0CECE}   \textbf{LOB}   & 4                                        & 4                                   & --                                      \\ \hline
\textbf{INF}   & 3                                        & 3                                   & 3                                       \\ \hline
\rowcolor[HTML]{D0CECE}   \textbf{MRD}   & 2                                        & 1                                   & --                                      \\ \hline
\textbf{INN}   & 1                                        & 0                                   & --                                      \\ \hline
\rowcolor[HTML]{D0CECE}   \textbf{IDS}   & 1                                        & 1                                   & 1                                       \\ \hline
\textbf{MNL}   & 1                                        & 1                                   & --                                      \\ \hline
\rowcolor[HTML]{D0CECE}   \textbf{IFL}   & 1                                        & 0                                   & --                                      \\ \hline
\textbf{IBS}   & 0                                        & 0                                   & --                                      \\ \hline
\rowcolor[HTML]{D0CECE}  \textbf{Other} & 12                                       & --                                  & --                                      \\ \hline
\end{tabular}
\label{tab:ablation}
\end{table}

\subsection{RQ2 (Ablation)}
\label{sec:ablation}
Table~\ref{tab:ablation} shows the performance of \ourapproach{} on different types of bugs found in DL programs in our benchmark. \textit{LLM} is the most prevalent bug type occurring in real-world buggy programs obtained from \sof in our benchmark. \ourapproach successfully detected 33/35 bugs in this category. Whereas, \nlint successfully detected 11/35 bugs of this category. For the second-most prevalent bug type \textit{CNL}, \ourapproach{} correctly identified 4/5 bugs, and, \nlint detected 1/5 bugs.
For bugs specific to CNN programs, \textit{INF} and \textit{IDS}, both \ourapproach and \nlint were able to detect all the bugs of these categories.
There are 12 bugs represented by the ``Other'' column in Table~\ref{tab:ablation} which are not supported by both \ourapproach and \nlint.
\ourapproach{} detected 57/75 bugs from different categories, while \nlint detects 17/75 bugs in these real-world buggy programs.

\subsection{RQ3 (Limitation)}
The scope of \ourapproach is defined as FCNN and CNN programs designed for regression and classification tasks using two deep-learning libraries, \Keras and \PyTorch.
Other architectures like Recurrent Neural Networks (RNNs) or pretrained DL models are not supported by \ourapproach. 
\ourapproach can be extended to support other architectures by adding new rules specific to those architectures.
\ourapproach is designed to detect 12 structural bugs; therefore, as shown in Table~\ref{tab:ablation} (Bug Categories - Other), it failed to find bugs due to insufficient data, wrong optimizer, incorrect weight initializer, epochs, and dropout rate.
As \ourapproach detects bugs at the beginning of the training, some of these bugs, \eg insufficient data cannot be detected before training.
Similarly, different optimizers have different convergence rates which cannot be determined at the early stage of training.
Identifying such bugs is a limitation of our approach.
Therefore, \ourapproach failed to detect 12 bugs in our benchmark.
We aim to address these bugs by integrating training monitoring into \ourapproach in the future.

\subsection{Result and Discussion}
Our technique, \ourapproach focuses on identifying structural defects, which are mainly caused by mistakes made by developers during the design of DL programs.
These design mistakes may have severe consequences which lead to incorrect output or poor generalization after training the DL model.
Detecting these flaws at an early stage of the training process has potential to save computational resources and the developer's time.
The results show that
\ourapproach outperforms state-of-the-art \nlint. 
Specifically, for real-world programs from \sof, \ourapproach identified and localized 34/45 bugs found in 22 buggy DL programs designed for multiclass classification tasks. However, \nlint detected 13/45 bugs in 22 buggy DL programs.
For binary classification and regression tasks, \ourapproach detected and localized 23/30 bugs found in 18 buggy DL programs, whereas, \nlint identified 4/30 bugs.
In total, \ourapproach successfully detected 57/75 bugs in 40 real-world buggy programs obtained from \sof. 
However, \ourapproach failed to detect 12 bugs from our benchmark as these bugs provide symptoms during training and \ourapproach does not support them. We plan to investigate these bugs in our future work.

\section{Threats to Validity}
\label{sec:threatstovalidity}

\textbf{External Threat:} We meticulously selected 105 posts from the dataset provided by \cite{humbatova20taxonomy} to understand the mapping between different types of bugs and dataset characteristics used in these posts to fix the bug. 
To enhance the generalizability of our research for future work, we propose incorporating additional sources, such as GitHub, to validate the applicability of the proposed approach across a broader range of real-world use cases.
Additionally, the design of our verification rules was influenced by insights from the literature~\cite{LeCun98LeNet5,Krizhevsky12AlexNet,Simonyan14VGG,backprop92,Bengio12,Baker17,LeCun89,Krizhevsky2010,Krizhevsky09cifar10,Shea15}, which could impact the generalizability of the study.
To mitigate potential biases, we utilize the defect4ML benchmark, which comprises 100 buggy DL programs collected from \sof and \gh, encompassing various bug categories. This benchmark serves as a reliable means to evaluate our proposed methodology.
We acknowledge that the conclusions drawn from this study provide an initial exploration of the bug categories and the challenges DL developers face in addressing these issues.
To evaluate our approach, we considered “Recommended
Fix from SO” as ground truth. To mitigate the bias due to the selection of the fixes as the ground truth, we applied the patch/fix to the buggy models and evaluated the model's performance before and after the fix.
Also, to mitigate the bias due to randomness in DNN models, we ran each program three times and compared the average accuracy of both the buggy and repaired programs.
We observed that these patches improved the 
performance of all the 40 models.

\textbf{Internal Threat:}
We were primarily concerned about the implementation of our verification rules.  Each rule requires to exact different layers of the model in sequence. To mitigate this threat, after designing and implementing \ourapproach, the authors carefully reviewed the code to reduce the chances of errors. 
We evaluate our approach, \ourapproach on 40 buggy DL programs. We considered \texttt{``Recommended Fix from SO''} as ground truth to evaluate our approach. As \ourapproach detects some bugs that were not specified in the ground truth, we need to verify the bugs reported are not false positive.
To mitigate this threat,
we verified the correctness of these bugs. Two authors independently examined the output generated by \ourapproach. They fixed the bugs using the actionable fixes reported by \ourapproach and checked the accuracy before and after the fix. If the model's performance is improved, the reported bugs are not considered false positives.

\section[Related]{Related Work}
\label{sec:relatedwork}
\subsection{Empirical study on Deep Learning Bugs}
In recent years, several empirical studies have investigated types of bugs in DL programs \cite{islam19,zhang18,humbatova20taxonomy}.
These studies have examined the symptoms and root causes of the deep learning bugs using the \sof posts and GitHub commits.
Zhang~\etal~\cite{zhang18} have studied the TensorFlow program bugs and identified 4 symptoms and 7 root causes for these bugs.
Meanwhile, Islam~\etal~\cite{islam19} studied real-world bugs in programs based on five deep-learning libraries 
\Caffe, \Keras, \tensor, \Theano, and \PyTorch, 
and identified 5 types of bugs and 10 root causes for these bugs.
They have also studied the impacts of these bugs on DL programs.
Another study was conducted by Islam~\etal~\cite{islam20repairing} to understand the bug fix patterns in DL programs and the challenges and risks involved in fixing them.
The study finds that bug localization and fixing is very difficult in DL programs as fixing one bug may introduce new bugs in the code.
Humbatova~\etal~\cite{humbatova20taxonomy} has provided a taxonomy of real faults in Deep Learning Systems. The faults are divided into 5 broad categories. Their study states that the faults in \texttt{Model} and \texttt{Training} categories mostly lead to performance-related issues.
whereas faults in the other three categories "GPU Usage", "API" and "Tensors and Inputs" leads to a crash.
Cao \etal~\cite{cao2022understanding} conducted the first comprehensive study to characterize performance problems in the Deep learning systems designed using TensorFlow and Keras. However, this work focused on the impact of time and resources (\eg GPU memory and power), whereas our work emphasizes localizing the structural bugs in FCNN and CNN models by analyzing the characteristics of datasets in real-world models written in \PyTorch and \Keras.

\subsection{Fault localization for Deep Learning Programs}

Due to the reliability on a lot of hyperparameters, the bugs in DL programs are different from the traditional software programs.
As the traditional fault localization techniques cannot be applied directly to DL programs which drew the researcher's attention to develop new techniques for fault localization in DL programs.
Therefore, various approaches have been proposed in the past for automatically detecting, localizing, and repairing DL program bugs. Nikanjam~\etal~\cite{nikanjam2021neuralint} proposed \nlint, a static analysis approach for automatic fault detection in deep learning programs. \nlint identifies the root cause of the bug based on pre-defined verification rules and also provides a message suggesting how to fix the bug. Although \nlint can detect bugs in FCNN models and is also capable of detecting bugs specific to CNN architecture, the goals of \nlint and \ourapproach are the same. However, \ourapproach considers the characteristics of the training dataset to automatically detect bugs, which allows it to outperform \nlint by detecting more structural bugs.
Schoop~\etal~\cite{schoop2021umlaut} proposed UMLAUT, which debugs DL programs using program structure and model behavior. 
Eniser \etal~\cite{eniser19deepfault} proposed DeepFault, which identifies suspicious neurons for fault localization in DL programs. 
Wardat \etal~\cite{wardat21DeepLocalize} propose DeepLocalize, a dynamic fault localization technique for DL programs.
DeepDiagnosis~\cite{wardat22DeepDiagnosis} is another dynamic fault localization technique that detects various symptoms during training and provides actionable fixes.
A learning-based fault diagnosis and localization approach DeepFD is proposed by Cao \etal~\cite{cao22deepfd} which maps fault localization tasks to a learning problem.
Braiek~\etal~\cite{BraiekDeepCheck} proposed a property-based debugging approach that detects bugs in three phases, \ie pre-training, during training, and post-fitting.
Although \cite{schoop2021umlaut,eniser19deepfault, wardat21DeepLocalize, wardat22DeepDiagnosis, cao22deepfd, BraiekDeepCheck} can detect bugs in DL programs, however, these approaches do not support CNN-architecture-specific bugs.
Ghanbari~\etal~\cite{ghanbari2023deepmufl} proposed a mutation-based fault localization approach for DL programs in which the mutants of pre-trained model are created to detect the bugs in DL programs.
Despite supporting faults related to CNN architecture, such as strides and filters in the convolution layer, it discovers bugs post-training. 
In contrast to these dynamic fault localization approaches, our approach, \ourapproach, works at the beginning of the training process and identifies the inappropriate configurations that results in faulty behavior during training.
This makes \ourapproach significantly faster than these approaches.

\section[Conclusion]{Conclusions and Future Work}
\label{sec:conclusion}
We propose an approach, named \ourapproach, to automatically detect 12 structural bugs in DL programs designed using two deep learning libraries, \Keras and \PyTorch.
We considered the characteristics of the training dataset and defined verification rules to localize them.
\ourapproach utilizes these rules to detect the bugs, localize them, and alert the developer with an informative message containing actionable fixes in buggy DL programs.
The bug's location and descriptive message help the developer easily locate the bug and improve the structure of the DL program.
\ourapproach performs bug localization at the beginning of the training process, thereby saving the time and computational resources of the developer.
\ourapproach outperforms state-of-the-art \nlint by localizing and suggesting the correct fixes for 57/75 buggy programs in our benchmark.
In the future, we plan to expand \ourapproach to support other architectures like RNNs.

\section[Data Availability]{Data Availability}
\label{sec:data-availability}

The benchmark consisting of 40 buggy DL programs obtained from \sof, files associated with our manual labeling process, and source code of \ourapproach are available in this 
repository~\cite{myRepo} which allows other researchers to reproduce the results for future research.

\section[Acknowledgments]{Acknowledgments}
\label{sec:Acknowledgment}

The authors thank the anonymous ICSE 2024 and TOSEM reviewers for their valuable feedback and Ali Ghanbari for his insightful comments.
This work is supported by the US National Science Foundation (NSF) under grants \#2512857, \#2512858, CCF-15-18897, CNS-15-13263, CNS-21-
20448, CCF-19-34884, CCF-22-23812, and NRT-21-52117, the Fonds de Recherche du Quebec (FRQ), the Canadian Institute for Advanced Research (CIFAR), and the Natural Sciences and Engineering Research Council of Canada (NSERC).
All opinions are those of the authors and do not reflect the views of the sponsors.

\balance

\bibliographystyle{ACM-Reference-Format}
\bibliography{paper}

\end{document}